\documentclass[14pt]{article}
\usepackage[utf8]{inputenc}
\usepackage{cite}
\usepackage{authblk}
\usepackage{graphicx}
\usepackage{subfig}
\usepackage[margin=1.1in]{geometry}
\usepackage{amsmath}
\usepackage{amssymb}
\usepackage{mathrsfs}
\usepackage{physics}
\usepackage{cancel}
\usepackage{comment}
\usepackage{tikz}
\usepackage{mwe}
\usepackage{mathtools}
\RequirePackage[colorlinks,citecolor=blue,urlcolor=magenta,linkcolor=blue]{hyperref}
\allowdisplaybreaks
\labelformat{section}{Section #1} 
\labelformat{subsection}{Section #1} 
\labelformat{subsubsection}{Section #1}
\labelformat{subsubsubsection}{Section #1}
\labelformat{equation}{Eq.~(#1)} 
\labelformat{subfigure}{Fig.~\thefigure#1} \labelformat{table}{Table~#1} 
\labelformat{appendix}{Appendix #1}

\usepackage{geometry}
\usepackage{hyperref}
\definecolor{darkslateblue}{rgb}{0.28, 0.24, 0.55}

\makeatletter
\def\@fnsymbol#1{\ensuremath{\ifcase#1\or \ddagger\or \dagger\or
   \mathsection\or \mathparagraph\or \|\or **\or \ddagger\ddagger
   \or \dagger\dagger \else\@ctrerr\fi}}

\title{ Zeroth law of black hole mechanics in general Horndeski theories} 

\author[1]{Ribhu Paul\footnote{ribhupaul.rp@gmail.com}}
\author[2]{Krishnakanta Bhattacharya\footnote{krishnakanta@dubai.bits-pilani.ac.in}}
\affil[1]{{\small{{{School of Physical Sciences \authorcr
Indian Association for the Cultivation of Science, Kolkata - 700032, India}}}}}
\affil[2]{{\small{{{Department of General Science \authorcr
BITS Pilani, Dubai Campus, International Academic City, Dubai, United Arab Emirates}}}}}

\date{}

\begin{document}

\maketitle

\begin{abstract}

We investigate the validity of the zeroth law in the general class of scalar–tensor theories with second-order field equations, namely Horndeski theories or generalized Galileon theories in four-dimensional spacetime. Assuming that the non-minimally coupled scalar field inherits the spacetime symmetries, we show that the horizon mixed projection of the field equations implies constant surface gravity on a stationary Killing horizon. Our analysis incorporates the most generic non-minimal scalar couplings permitted within the Horndeski framework, which we consider as a perturbative expansion about general relativity. Our results therefore classify the possible cases within this framework of proof for different classes of scalar-tensor theories, including shift-symmetric scalar fields. These results extend the applicability of black hole mechanics beyond Einstein gravity and provide further evidence for the thermodynamic consistency of general Horndeski theories.
\end{abstract}

\section{Introduction}
A notable correspondence exists between black hole mechanics and standard thermodynamics, summarized by the four laws\cite{Bardeen:1973gs}. Analogous to the zeroth law in thermodynamics, which defines the notion of temperature at thermal equilibrium, the zeroth law of black hole mechanics states the constancy of surface gravity on the stationary Killing horizon. This analogy is reinforced by the relationship between surface gravity and the temperature of a stationary black hole's horizon, in the context of Hawking radiation\cite{Hawking:1975vcx}, thereby establishing a classical thermodynamic correspondence. Various proofs of the black hole zeroth law exist in the literature, each with its unique applicability of the theory in hand. Further, in the context of stationary spacetimes, the rigidity theorem \cite{Hawking:1973uf, Hawking:1972qk} ensures that an event horizon is also a Killing horizon, allowing surface gravity to be defined in terms of the stationary Killing vector that behaves as a null vector on the horizon. This theorem, which is essential for proving the zeroth law, was further extended to diffeomorphism-invariant theories by treating them as effective field theories\cite{Hollands:2022ajj}. Additionally, since general relativity (GR) is known to be UV incomplete, it can naturally be extended within a low-energy effective field theory framework by incorporating higher-order corrections and additional degrees of freedom. In such modified theories of gravity, however, there exists no proof for the zeroth law in general, as it depends on the effective equations of motion and suitable assumptions\cite{Sang:2021rla}. Quite notably, the zeroth law was proven in a theory-agnostic manner, considering very strong restrictions ($t-\varphi$ reflection isometry) on the nature of the black hole solutions\cite{Racz:1995nh}. \par
Beyond the theoretical considerations, there are also observational motivations for exploring extensions of GR. Although GR has been remarkably successful in describing gravitational phenomena, observations on cosmological scales point to regimes where the theory may require further extension\cite{SupernovaCosmologyProject:1998vns, SupernovaSearchTeam:1998fmf}. Major cosmological puzzles, such as the nature of dark matter, dark energy, and the origins and evolution of the early universe, remain unanswered within this framework. To explain the late-time accelerated expansion of the universe, modified gravity theories have been widely explored, in particular, ones featuring dynamical scalar field models of dark energy which can be strongly constrained from observations\cite{Kase:2018aps}. Thus, one potential extension to GR involves incorporating a dynamical scalar field into the prevalent metric theory, which has been widely studied\cite{Callan:1985ia, Esposito-Farese:2003jfi, Briscese:2006xu, Saridakis:2016ahq, Haridasu:2021hzq, Lazanu:2024mzj}.\par
Scalar-tensor (ST) theories such as Brans-Dicke theory\cite{Brans:1961sx, Fujii:2003pa} introduce a non-minimally coupled scalar field $\phi$, alongside the metric $g$, as a modification to general relativity, while others feature a minimally coupled dynamical scalar field\cite{Armendariz-Picon:2000ulo}. These theories, including those with non-canonical kinetic terms and nonminimal coupling to curvature, have been used in models of inflation and dark energy (cf. for details--\cite{Sotiriou:2008rp, CANTATA:2021ktz, Armendariz-Picon:1999hyi}). The introduction of additional scalar degrees of freedom naturally arises in a variety of settings, including effective field theories, cosmological models of inflation and dark energy, higher-dimensional theories, and low-energy limits of string-inspired models. As a general extension to scalar-tensor theories, the Horndeski theories \cite{Horndeski:1974wa, Kobayashi:2011nu, Deffayet:2009wt}, provide the most generic structure for ST theories with second-order equations of motion, devoid of Ostrogradskii instability. The original Lagrangian for Horndeski theories was further shown to be equivalent to generalized Galileon theories\cite{Kobayashi:2019hrl} via a specific map, which enables them to be considered on equal terms. These theories are characterized by arbitrary and regular coupling functions $\mathcal{G}_{2,3,4,5}(\phi, X)$, where $X$ denotes the canonical kinetic term, and are crucial for exploring deviations from GR in strong-field regimes, particularly through the properties of black hole solutions\cite{Minamitsuji:2023nvh}. There has been interest in the thermodynamic properties of black holes in the context of Horndeski gravity\cite{Minamitsuji:2023nvh}. Moreover, for Brans-Dicke-like theories that are conformally related to GR with minimally coupled matter, and where the parameter\footnote{known as the Brans-Dicke parameter.} $\omega_{\text{BD}}\to \omega(\phi)$, extensive studies\cite{Bhattacharya:2017pqc, Bhattacharya:2020wdl, Dey:2021rke} have revealed the understanding of thermodynamic connections and black hole mechanics.\par
While the first law of black hole mechanics and the associated notion of Wald entropy\cite{Wald:1993nt} have been extensively studied in diffeomorphism-invariant theories, comparatively less attention has been devoted to the zeroth law in general scalar-tensor theories. In GR, Einstein's field equations\cite{Bardeen:1973gs}, along with the dominant energy condition (DEC)\footnote{Without further restrictions imposed on the geometry of the horizon.}, can demonstrate the constancy of surface gravity. It therefore does not immediately generalize to theories containing non-minimal scalar-curvature couplings. Indeed, such couplings introduce additional geometric and scalar contributions to the field equations that might hamper the conventional argument leading to the constancy of surface gravity. It is therefore far from obvious that the zeroth law should remain valid in the full Horndeski class. Using a perturbative expansion of the coupling constants, and assuming GR as a smooth limit, the zeroth law has been established in higher curvature gravity\cite{Ghosh:2020dkk}, and later in arbitrary higher derivative theories of gravity\cite{Bhattacharyya:2022nqa}.\par
This paper thus aims to prove the zeroth law in the generic framework of Horndeski theories (generalized Galileon theories in four-dimensional spacetime). However, a further symmetry restriction on the nature of the scalar field  $\phi$ is required, i.e., the Killing symmetry condition holds for $\phi$, as there may be solutions where the above may not be true (for example, cf.---\cite{Babichev:2013cya,Kobayashi:2014eva,Babichev:2016rlq}). In the latter part, we uplift the restriction and analyze its consequence. Our proof treats general relativity as a smooth limiting theory and therefore extends previous perturbative analyses performed for higher-curvature and higher-derivative gravitational theories\cite{Ghosh:2020dkk,Bhattacharyya:2022nqa}. The results provide further evidence that the thermodynamic structure underlying black hole mechanics survives within a large class of physically relevant scalar-tensor theories.\par The paper is structured as follows: First, we begin by reviewing some geometric properties of stationary Killing horizons and stating preliminary concepts required for further analysis (\ref{sec:stationary}). Next, we introduce the generalized Horndeski theories studied here and provide an overview of the corresponding equations of motion and geometric quantities (\ref{sec:horndeski}). Following that, we establish the proof of the zeroth law (\ref{sec:zeroth_law}) for stationary Killing horizons in the context of these generalized Horndeski theories. Lastly, we conclude with a summary of the results obtained (\ref{sec:conclusion}) and potential implications.\par
\textit{Notations \& Conventions}: We consider four-dimensional spacetime $(\mathscr{M}, g)$ endowed with an additional scalar field $\phi$, and a compatible metric with mostly positive signature \((- , + , + , +)\). We also use geometrized units where \( c=\hbar=16\pi G=1\). Lowercase Latin initial letters \( a, b, \ldots \) indicate spacetime indices, ranging from 0 to 3, while the uppercase Latin initial letters \( A, B, \ldots \) represent the transverse coordinates of the two-dimensional cross-section $\mathscr{C}$ of the null hypersurface $\mathcal{H}\subset\mathscr{M}$, ranging from 2 to 3. Additionally, for notational convenience, the $n-$th order 4-covariant derivative of any smooth scalar function $f$ on $\mathscr{M}$ i.e. $\nabla_{a_1} \nabla_{a_2}\ldots\nabla_{a_n}f$ shall be denoted as $f_{a_1a_2\ldots a_n}$: for example, $\nabla_a\phi\equiv \phi_a$ ;~$\nabla_{a}\nabla_{b}\phi\equiv \phi_{ab}$ and so on. Tranverse/induced covariant derivative on $\mathscr{C}$ is denoted as $D_aS_{bcd\ldots}=h_a^mh^n_bh^p_ch^q_d \ldots\nabla_m S_{npq\ldots}$, where $h_{mn}$ is the induced metric on the cross-section. Lastly, parentheses enclosing a group of indices signify that the indices are symmetrized, and each is assigned a weight of one. Therefore, for a rank-$n$ covariant tensor $T_{a_1\ldots a_n}$, we can express:
$\displaystyle{
T_{(a_1\dots a_n)} \equiv \frac{1}{n!} \sum_{\sigma \in S_n} T_{a_{\sigma(1)}\dots a_{\sigma(n)}}}$,
where $S_n$ represents the group of all the possible permutations of $n$ indices. For two indices, the symmetrization convention reduces to
$\displaystyle{T_{(ab)}
\equiv
\frac{1}{2}\left(T_{ab}+T_{ba}\right)}$.
\section{Geometric preliminaries of stationary Killing horizon}\label{sec:stationary}
A central role in the proof of the zeroth law is played by the geometry of stationary Killing horizons. We consider $\mathcal H$ to be a null hypersurface generated by a timelike Killing vector field $\xi^a$ that becomes null on $\mathcal H$, or
\begin{equation*}
    \xi^a\xi_a \overset{\mathcal{H}}{=} 0~.
\end{equation*}
Since the norm of $\xi^a$ vanishes on the horizon, the Killing vector is simultaneously tangent and normal to $\mathcal H$, and its integral curves define the null generators of the horizon. The failure of these generators to be affinely parametrized is characterized by the surface gravity $\kappa$, defined through
\begin{equation}\label{kappa_def}
    \xi^a\nabla_a\xi^b =\kappa \xi^b~. \qquad 
\end{equation}
The physical weight of $\kappa$ comes from its role in Hawking's semiclassical result \cite{Hawking:1975vcx}. This is because a stationary black hole radiates thermally at temperature $T_{\rm{H}}=\kappa/2\pi$, and this identification is what elevates $\kappa$ from a purely geometric quantity (the inaffinity of the horizon generators) to a thermodynamic one. It is, however, far from self-evident that a single number $T_{\rm{H}}$ should suffice to describe the entire horizon, as $\kappa$ is, a priori, a function on $\mathcal{H}$  (free to vary from one point to the next on $\mathcal{H}$). The zeroth law is precisely the statement that this does not happen, i.e. $\kappa$ takes one constant value over the whole stationary Killing horizon, exactly as required for $T_{\rm{H}}$ to be a well-defined equilibrium temperature in the first place. This is the physical content we are after; the remainder of this section only sets up the geometric language needed to state it precisely.

In our subsequent analysis, we shall choose a suitable basis $\{\xi^a, v^a, \mathbf{e}^a_{A=1,2}\}$, where $\mathbf{e}^a_A$ are tangent vectors of the smoothly connected horizon cross-section $\mathscr{C}$, satisfying the following relations
\begin{equation}
    \xi^a v_a = -1, \quad v^a v_a = 0, \quad \xi_a \mathbf{e}^a_A = 0, \quad v_a \mathbf{e}^a_A = 0, \quad g_{ab} \mathbf{e}^a_A \mathbf{e}^b_B = \delta_{AB}~.
\end{equation}
Therefore, in this basis we have
\begin{equation*}
    \kappa =-v_b\xi^a\nabla_a\xi^b~.
\end{equation*}
Lastly, the horizon metric is given as 
\begin{equation*}
    g_{ab}=-2\xi_{(a}v_{b)}+h_{ab}~,
\end{equation*}
The cross-section induced metric is denoted by $h_{ab}$ and applies to the spatial cross section $\mathscr{C}$. At each point $p$ on the Killing horizon $\mathcal{H}$, the tangent space $T_p\mathcal{H}$ is three-dimensional. If a spacelike two-dimensional cross-section $\mathscr{C}$ of $\mathcal{H}$ is chosen, then the tangent space can be decomposed as
\begin{equation}
    T_p\mathcal{H}=T_p\mathscr{C}\oplus\mathrm{span}\{\xi^a\}~.
\end{equation}
Thus, variations of any scalar defined on the horizon can be separated into variations along the null generators, generated by $\xi^a$, and variations across the generators, tangent to the cross-section $\mathscr{C}$. Consequently, to establish that the surface gravity $\kappa$ is constant over the Killing horizon $\mathcal{H}$, it is sufficient to show that $\xi^a\nabla_a\kappa=0~,$
and that its gradient has no component tangent to the spacelike cross-section $\mathscr{C}$, namely, $h_a^b\nabla_b\kappa=0~,$
where $h^a_b$ denotes the projection tensor onto $\mathscr{C}$. The first condition follows directly from the Killing symmetry and \ref{kappa_def}, therefore expressing the constancy of $\kappa$ along the null generators of $\mathcal{H}$. The second condition, which establishes the absence of variation of $\kappa$ across the generators, constitutes the non-trivial part of the proof. Together, these conditions imply that $\kappa$ is constant over the entire Killing horizon,
\begin{equation}
    \kappa\overset{\mathcal{H}}{=}\mathrm{constant}~.
\end{equation} 
As a first step towards proving the zeroth law, one thus needs to show the constancy of $\kappa$ on the cross-section $\mathscr{C}$, which amounts to showing that the transverse covariant derivative of $\kappa$ vanishes, i.e. $D_a \kappa =0$. As we shall see, the equation of motion for a general scalar-tensor theory when contracted with $\xi^a\xi^b$ takes the form
\begin{equation}
    \mathfrak{H}_{\xi\xi}\equiv\mathfrak{H}_{ab}\xi^a\xi^b=\mathscr{F}(\mathcal{L}_{\xi}\phi, \mathcal{L}^2_{\xi}\phi; \mathscr{G})~,
\end{equation}
with $\mathfrak{H}_{ab}$ representing the modified Einstein tensor, which in fact becomes $G_{ab}=R_{ab}-\frac{1}{2}g_{ab}R$ in GR.
{Here, it is to be mentioned that $\mathscr{G}$ is dependent on certain geometrical facts that can be laid down as follows}
\begin{equation}\label{geom_identities}
R_{\xi\xi}\equiv R_{ab}\xi^a\xi^b=0~;~R_{acbd}\xi^a\xi^b\mathbf{e}^{c}_A\mathbf{e}^{d}_C=0~;~R_{aecd}\xi^a\mathbf{e}^e_A\mathbf{e}^c_C\mathbf{e}^d_E=0~,
\end{equation}
The proof of the identity $\mathfrak{H}_{\xi\xi}=0$ repeatedly relies on the above set of geometric identities valid on a stationary Killing horizon. Since the horizon is generated by a Killing field, the induced metric on any spatial cross-section is Lie dragged along the generators, implying that both the expansion and shear vanish identically. Raychaudhuri's equation then yields the first identity \ref{geom_identities}. Furthermore, the Gauss-Codazzi \cite{Vega:2011ue} and Ricci embedding equations imply that curvature components containing two horizon generators or one horizon generator together with three transverse directions vanish, implying the other two identities \ref{geom_identities}.
It is important to note here that the functional form $\mathscr{F}(\pounds_{\xi}\phi, \pounds^2_{\xi}\phi; \mathscr{G})$ is polynomial in its arguments and vanishes completely only when $\pounds_{\xi} \phi=0$. Thus, it immediately follows that $\mathscr{F}(\pounds_{\xi}\phi, \pounds^2_{\xi}\phi; \mathscr{G})=0=\mathfrak{T}_{ab}\xi^a\xi^b$ on $\mathcal{H}$, in the presence of additional minimally coupled matter action $\mathcal{A}_{\rm mat}[g_{ab},\Psi]$ with $\Psi:=\{\psi_1,\psi_2,\ldots\}$ representing the minimally coupled matter fields in general. This then forms a standard procedure, thereby implying that $\mathfrak{T}_{ac}\xi^ah^c_b=0$, provided the matter satisfies the dominant energy condition. The dominant energy condition (DEC) is the statement that, for any observer, the local energy density is non-negative and the energy flux never propagates faster than light; equivalently, $-\mathfrak{T}^a_{~b}\xi^b$ is a future-directed causal vector for every future-directed causal $\xi^b$. It is the condition one expects to hold for essentially all physically reasonable matter -- ordinary fluids, fields, and radiation -- short of genuinely exotic sources, and it is exactly this input, rather than any detail of the matter's internal dynamics, that converts the vanishing of $\mathfrak{T}_{ab}\xi^a\xi^b$ into the stronger statement $\mathfrak{T}_{ac}\xi^ah^c_b=0$ used throughout. While this is a step closer towards proving the zeroth law (which is obvious for GR), one needs to further assume the smooth limit of all physical quantities to GR to prove in full generality\cite{Ghosh:2020dkk}.
In the next section, we introduce and discuss general Horndeski theories, their general structure and associated field equations that will be relevant in our subsequent proof.
\section{General Horndeski theories (DGSZ action)}\label{sec:horndeski}
We begin with the DGSZ (Deffayet--Gao--Steer--Zahariade) formulation of Horndeski theories \cite{Deffayet:2011gz}, which is equivalent to the original construction proposed by Horndeski \cite{Horndeski:1974wa}. Horndeski theories occupy a distinguished place among scalar-tensor extensions of GR: they comprise the most general four-dimensional theory of a metric $g_{ab}$ non-minimally coupled to a single scalar $\phi$ whose field equations remain second order, thereby avoiding the Ostrogradski instability that would otherwise plague a generic higher-derivative scalar-tensor theory \cite{Woodard:2015zca}. Rather than committing to any one extension of GR at the outset, the theory is organized around four arbitrary coupling functions $\mathcal{G}_{2,3,4,5}(\phi,X)$, each of which, by itself, reproduces a familiar and well-studied model. The total action (along with matter) can be written as
\begin{equation}\label{full_action}
\begin{aligned}
\mathcal{A}[\phi,g_{ab};\Psi]
&=
\mathcal{A}_{\rm H}[\phi,g_{ab}]
+
\mathcal{A}_{\rm mat}[g_{ab},\Psi]\\
&=
\int_{\mathscr{M}} d^4x,\sqrt{-g}
\left[
k(\phi,X)
+\sum_{i=3}^{5}\mathcal{L}_{\rm H}^{(i)}(g_{ab},\phi)
\right]
+
\int_{\mathscr{M}} d^4x,\sqrt{-g},
\mathcal{L}_{\rm mat}(g_{ab},\Psi)~,
\end{aligned}
\end{equation}
where $k(\phi,X)$ often denoted by $\mathcal{G}
_2(\phi, X)$ is the $k$-essence sector, which generalizes the canonical scalar-field kinetic term and encompasses the class of $k$-essence models commonly considered in cosmology. The remaining Horndeski Lagrangians are given by
\begin{equation}
\begin{aligned}
\mathcal{L}_{\rm H}^{(3)}&=-\mathcal{G}_3\Box\phi,\
\mathcal{L}_{\rm H}^{(4)}&=\left(
\mathcal{G}_4 R^{ab}_{cd}
+2\mathcal{G}_{4X}\phi^a_{c}\phi^b_{d}\right)
\delta_{ab}^{cd},\
\mathcal{L}_{\rm H}^{(5)}&=\mathcal{G}_5G^a_{b}\phi^b_{a}+\mathcal{G}_{5X}\delta^{pqr}_{abc}\phi^a_{p}\phi^b_{q}\phi^c_{r}.
\end{aligned}
\end{equation}
For compactness, we have defined the generalized antisymmetric Kronecker delta with $2p$ indices as
\begin{equation}
    \delta^{a_1 \dots a_p }_{b_1 \dots b_p} \equiv\frac{(-1)^p}{p_!}
\begin{vmatrix}
\delta^{a_1}_{b_1} & \cdots & \delta^{a_1}_{b_p}~ \\
\vdots & \ddots & \vdots \\
\delta^{a_p}_{b_1} & \cdots & \delta^{a_p}_{b_p}~
\end{vmatrix}~,
\end{equation}
where all indices take values in ${0,1,2,3}$.
It is to be noted that the $\mathcal{G}_3\Box\phi$ sector corresponds to the generalized cubic Galileon and constitutes the kinetic-gravity-braiding sector \cite{Deffayet:2010qz}. The matter Lagrangian $\mathcal{L}_{\rm mat}(g_{ab},\Psi)$ describes the external matter degrees of freedom $\Psi$ and is minimally coupled to the spacetime metric $g_{ab}$, which is also identified as the Jordan-frame metric.
The Horndeski functions $\mathcal{G}_i(\phi, X)$, for $i=2,3,4,5$, are arbitrary functions of the scalar field and its canonical kinetic term,
\begin{equation}
X\equiv-\frac{1}{2}g^{cd}\phi_c\phi_d.
\end{equation}
We adopt the standard notation
\begin{equation}
\mathcal{G}_{iX}\equiv\frac{\partial\mathcal{G}_i}{\partial X}~,
\quad
\mathcal{G}_{i\phi}\equiv\frac{\partial\mathcal{G}_i}{\partial\phi}~,
\end{equation}
together with
\begin{equation}
\phi^c_{d}\equiv g^{ce}\phi_{ed}~,
\quad
\phi^{cd}\equiv g^{de}\phi^c_{e}~.
\end{equation}

The $\mathcal{G}_4$ as well as $\mathcal{G}_5$ sector contains the non-minimal coupling $\mathcal{G}_4(\phi, X)R$ and $\mathcal{G}_5(\phi, X)G^a_b\phi^b_a$, thereby directly coupling the scalar sector to the curvature of spacetime. Hence, the gravitational dynamics are determined by the coupled scalar and tensor sectors rather than by the metric alone. Unlike Brans--Dicke theories, for which a conformal transformation can be performed to cast the theory into an Einstein-frame representation, a generic Horndeski theory does not, in general, admit a transformation to an Einstein frame in which the gravitational sector is reduced to the minimally coupled Einstein--Hilbert form.

On a well-posed variation of the full action \ref{full_action}, one obtains the following
\begin{equation}
\begin{aligned}
    \delta\mathcal{A}&=\int_{\mathscr{M}}d^4x\sqrt{-g}\left(\left(-\mathfrak{H}^{ab}+\frac{1}{2}\mathfrak{T}^{ab}\right)\delta g_{ab}+\mathcal{E}_{(\phi)}\delta\phi\right)+\int_{\partial\mathscr{M}}d^3x\sqrt{h} ~(P^{ab}\delta h_{ab}+P_{(\phi)}\delta\phi)\\&~~~-\delta\mathscr{B}_{PS}[h_{ab},\phi,u^c\nabla_ch_{ab},\mathcal{L}_u\phi]~,
\end{aligned}
\end{equation}
where, $\mathscr{B}_{PS}[h_{ab},\phi,u^c\nabla_ch_{ab},\mathcal{L}_u\phi]$ is the boundary term\cite{Padilla:2012ze}, such that $\mathscr{B}_{PS}[h_{ab},\phi,u^c\nabla_ch_{ab},\mathcal{L}_u\phi]\to 0$ as $u^c\nabla_ch_{ab}\to 0$ and $\mathcal{L}_u\phi\to 0$ simultaneously ($u$ is a timelike vector field).\\
Thus, in the presence of minimally coupled matter, we can infer the following field equations after fixing $h_{ab}$ and $\phi$ on $\partial \mathscr{M}$
\begin{equation}\label{eom}
    \mathfrak{H}^{ab}=\frac{1}{2}\mathfrak{T}^{ab}~;~\mathcal{E}_{(\phi)}=0~.
\end{equation}
Here, one can rewrite the lhs of the above tensor field equations\cite{Gao:2011mz} as 
\begin{equation}\label{expanded_eom}\mathfrak{H}_{ab}=\mathfrak{H}^{(1)}G_{ab}+\mathfrak{H}^{(2)}g_{ab}+\mathfrak{H}^{(3)}\phi_{a}\phi_{b}+\mathfrak{H}^{(4)}\phi_{ab}+\mathfrak{H}^{(5)}_{ab}~,\end{equation}
with 
\begin{equation}\label{group}
    \begin{aligned}
    \mathfrak{H}^{(1)}&=\mathcal{G}_4+\frac{1}{2}\nabla_c\mathcal{G}_5\phi^c~,\\
    \mathfrak{H}^{(2)}&=-\frac{1}{2}\mathcal{G}_2-\frac{1}{2}\nabla_{c}\mathcal{G}_3\phi^c+\mathcal{G}_{4\phi}\Box \phi-2X\mathcal{G}_{4\phi\phi}-\mathcal{G}_{4X}R^{cd}\phi_c\phi_d+\nabla_{c}\mathcal{G}_{4X}\phi^c\Box \phi-2\mathcal{G}_{4\phi X}\phi^{cd}\phi_{c}\phi_{d}\\
    &~~+\mathcal{G}_{4XX}\phi_{cd}\phi_{e}^{d}\phi^{c}\phi^{e}+\mathcal{G}_{4X}\phi^a_c\phi^d_d\delta^{cd}_{ab}+\mathcal{G}_{5X}\delta^{pqr}_{abc}\phi^a_q\phi^b_q\phi^c_r+\nabla_{c}\mathcal{G}_5R^{cd}\phi_d-\frac{1}{2}\nabla_{c}\left(\mathcal{G}_{5\phi}\phi^c\right)\Box\phi\\&~~+\frac{1}{2}\nabla_{c}\left(\mathcal{G}_{5\phi}\phi_{d}\right)\phi^{cd}-\frac{1}{2}\nabla_{c}\mathcal{G}_{5X} X^c\Box \phi+\frac{1}{2}\nabla_c\mathcal{G}_{5X}X_d\phi^{cd}+\frac{1}{2}\mathcal{G}_{5X}R_{cd}\phi^c\phi^d\Box\phi\\&~~-\frac{1}{2}\nabla^c\mathcal{G}_{5X}(\nabla_{c}\phi)\delta^{pq}_{ab}\phi^a_p\phi^b_q-\frac{1}{2}\mathcal{G}_{5X}R_{cdef}\phi^{ce}\phi^d\phi^f~,\\
    \mathfrak{H}^{(3)}&=-\frac{1}{2}\mathcal{G}_{2X}+\frac{1}{2}\mathcal{G}_{3X}\Box \phi-\frac{1}{2}\mathcal{G}_{4X}R-\mathcal{G}_{4XX}\delta^{cd}_{ab}\phi^a_c\phi^b_d-\mathcal{G}_{4\phi\phi}-\frac{1}{2}\mathcal{G}_{5X}G_{cd}\phi^{cd}-\frac{1}{2}\mathcal{G}_{5XX}\delta^{pqr}_{abc}\phi^a_q\phi^b_q\phi^c_r~,\\
    \mathfrak{H}^{(4)}&=-\mathcal{G}_{4X}\Box \phi-\nabla_{c}\mathcal{G}_{4X}\phi^c-\mathcal{G}_{4\phi}-\frac{1}{2}\mathcal{G}_{5X}R_{cd}\phi^c\phi^d+\frac{1}{2}\left[\nabla_{c}\left(\mathcal{G}_{5\phi}\phi^c\right)-\nabla_{c}\left(\mathcal{G}_{5\phi}\phi_d\right)\phi^c\phi^d\right]+\frac{1}{2}\phi^c\nabla_{c}\mathcal{G}_{5X}\Box \phi\\
    &~~+\frac{1}{2}\mathcal{G}_{5X}(\Box \phi)^2~,\\
    \end{aligned}
\end{equation}
and,
\begin{equation}
    \begin{aligned}     \mathfrak{H}^{(5)}_{ab}&=\nabla_{(a}\mathcal{G}_3\phi_{b)}+\mathcal{G}_{4X}\phi_{ca}\phi^{c}_{b}+2\nabla_{c}\mathcal{G}_{4X}\phi^{c}_{~(a}\phi_{b)}+2\mathcal{G}_{4X}R_{c(a}\phi_{b)}\phi^c-2\nabla_{(a}\mathcal{G}_{4X}\phi_{b)}\Box \phi\\
    &~~+\mathcal{G}_{4X}R_{acbd}\phi^c\phi^d+2\mathcal{G}_{4\phi X}\phi^{c}\phi_{c(a}\phi_{b)}-\mathcal{G}_{4XX}\phi^c\phi_{ca}\phi^d\phi_{db}+\mathcal{G}_{5X}R_{cd}\phi^{c}\phi^d_{(a}\phi_{b)}\\&~~-\mathcal{G}_{5X}R_{c(a}\phi_{b)}\phi^c\Box \phi-\frac{1}{2}\mathcal{G}_{5X}R_{acbd}\phi^c\phi^d\Box \phi+\mathcal{G}_{5X}R_{cde(a}\phi_{b)}\phi^d\phi^{ce}+\mathcal{G}_{5X}R_{cde(a}\phi^{d}_{b)}\phi^c\phi^e\\
    &~~-\frac{1}{2}\nabla_{(a}\left[\mathcal{G}_{5X}\phi^c\right]\phi_{cb)}\Box\phi+\frac{1}{2}\nabla_{(a}\left[\mathcal{G}_{5\phi}\phi_{b)}\right]\Box\phi-\nabla_{c}\left[\mathcal{G}_{5\phi}\phi_{(a}\right]\phi_{b)}^{c}+\nabla^{c}\mathcal{G}_5\phi^dR_{c(ab)d}-\nabla_{(a}\mathcal{G}_5 G_{b)c}\phi^c\\&~~+\nabla_{(a}\mathcal{G}_{5X}\phi_{b)}\delta^{cd}_{pq}\phi^p_c\phi^d_q-\frac{1}{2}\mathcal{G}_{5X}\Box \phi\phi_{ca}\phi^{c}_{b}-\frac{1}{2}\phi^c\nabla_c\mathcal{G}_{5X}\phi_{da}\phi^d_{b}-\nabla^c\mathcal{G}_5R_{c(a}\phi_{b)}\\&~~+\nabla_c\left[\mathcal{G}_{5X}\phi_d\right]\phi^{c}_{(a}\phi^{d}_{b)}-\nabla_d\mathcal{G}_{5X}\Box \phi \phi^{d}_{(a}\phi_{b)}+\nabla_d\mathcal{G}_{5X}\phi^{cd}\phi_{c(a}\phi_{b)}~.
    \end{aligned}
\end{equation}
The tensor and scalar sectors are connected by the Bianchi-like identity $\displaystyle{\nabla_a\mathfrak{H}^{a}_{b}+\frac{1}{2}\mathcal{E}_{(\phi)}\phi_b=0}~,$
which in turn implies that
\begin{equation}
    \nabla_a\mathfrak{T}^{a}_b=-\mathcal{E}_{(\phi)}\phi_b=0\rm ~~~(scalar ~onshell)~.
\end{equation}
Hence, the scalar equation along with diffeomorphism invariance and the metric equation implies matter conservation.
After having reviewed the general Horndeski framework, we now turn to the proof of the zeroth law.
\section{Proving the zeroth law}\label{sec:zeroth_law}
In this section, we investigate the zeroth law in general Horndeski gravity by dividing the calculation into three parts: first, contracting the equations of motion with \( \xi^a \xi^b \); second, contracting them with \( \xi^a h^b_c \); and third, analyzing the resulting identity and discussing it.
\subsection{Null-null projection: $\mathfrak{H}_{\xi\xi}$}\label{null_null_proj}
We first consider the null-null projection of the equations of motion \ref{eom}, which is obtained by contracting the lhs of the equation \ref{expanded_eom} with $\xi^a\xi^b$ 
\begin{equation}
    \mathfrak{H}_{\xi\xi}=\mathfrak{H}^{(1)}R_{ab}\xi^a\xi^b+\mathfrak{H}^{(3)}\phi_a\phi_b\xi^a\xi^b+\mathfrak{H}^{(4)}\phi_{ab}\xi^a\xi^b+\mathfrak{H}^{(5)}_{ab}\xi^a\xi^b~.
\end{equation}
We evaluate the above expression on the horizon.
Thus, the first term in the above expansion vanishes due to the relation \ref{geom_identities} on the horizon $R_{ab}\xi^a\xi^b=0$, while the second term vanishes with the assumption that the special solution of the non-minimally coupled scalar field $\phi$ has the same symmetries as spacetime: $\pounds_{\xi}\phi=0$. The third term also vanishes since
\begin{equation}\label{rel1}
\phi_{ab}\xi^a\xi^b\overset{\mathcal{H}}{=}\xi^a\nabla_a\left(\xi^b\phi_b\right)-\kappa \xi^b\phi_b=0~.
\end{equation}
We will use the above relation in future algebraic steps.
Thus, the only null-null projection we need to consider is the expression $\mathfrak{H}^{(5)}_{\xi\xi}\equiv\mathfrak{H}^{(5)}_{ab}\xi^a\xi^b$, which turns out to be
\begin{equation}
    \begin{aligned}
        \mathfrak{H}^{(5)}_{\xi\xi}&=\nabla_{a}\mathcal{G}_3\phi_{b}\xi^a\xi^b+\mathcal{G}_{4X}\phi_{ca}\phi^{c}_{b}\xi^a\xi^b+2\nabla_{c}\mathcal{G}_{4X}\phi^{c}_{a}\phi_{b}\xi^a\xi^b+2\mathcal{G}_{4X}R_{ca}\phi_{b}\phi^c\xi^a\xi^b-2\nabla_{a}\mathcal{G}_{4X}\phi_{b}\Box \phi\xi^a\xi^b\\
    &~~+\mathcal{G}_{4X}R_{acbd}\phi^c\phi^d\xi^a\xi^b+2\mathcal{G}_{4\phi X}\phi^{c}\phi_{ca}\phi_{b}\xi^a\xi^b-\mathcal{G}_{4XX}\phi^c\phi_{ca}\phi^d\phi_{db}\xi^a\xi^b+\mathcal{G}_{5X}R_{cd}\phi^{c}\phi^d_{a}\phi_{b}\xi^a\xi^b\\&~~-\mathcal{G}_{5X}R_{ca}\phi_{b}\phi^c\Box \phi\xi^a\xi^b-\frac{1}{2}\mathcal{G}_{5X}R_{acbd}\phi^c\phi^d\Box \phi\xi^a\xi^b+\mathcal{G}_{5X}R_{cdea}\phi_{b}\phi^d\phi^{ce}\xi^a\xi^b+\mathcal{G}_{5X}R_{cdea}\phi^{d}_{b}\phi^c\phi^e\xi^a\xi^b\\
    &~~-\frac{1}{2}\nabla_{a}\left[\mathcal{G}_{5X}\phi^c\right]\phi_{cb}\Box\phi\xi^a\xi^b+\frac{1}{2}\nabla_{a}\left[\mathcal{G}_{5\phi}\phi_{b}\right]\Box\phi\xi^a\xi^b-\nabla_{c}\left[\mathcal{G}_{5\phi}\phi_{a}\right]\phi_{b}^{c}\xi^a\xi^b+\nabla^{c}\mathcal{G}_5\phi^{d}R_{cabd}\xi^a\xi^b\\&~~-\nabla_{a}\mathcal{G}_5 G_{bc}\phi^c\xi^a\xi^b+\nabla_{a}\mathcal{G}_{5X}\phi_{b}\delta^{ef}_{cd}\phi^c_e\phi^d_f\xi^a\xi^b-\frac{1}{2}\mathcal{G}_{5X}\Box \phi\phi_{ca}\phi^{c}_{b}\xi^a\xi^b-\frac{1}{2}\phi^c\nabla_c\mathcal{G}_{5X}\phi_{da}\phi^d_{b}\xi^a\xi^b\\&~~-\nabla^c\mathcal{G}_5R_{ca}\phi_{b}\xi^a\xi^b+\nabla_c\left[\mathcal{G}_{5X}\phi_d\right]\phi^{c}_{a}\phi^{d}_{b}\xi^a\xi^b-\nabla_d\mathcal{G}_{5X}\Box \phi \phi^{d}_{a}\phi_{b}\xi^a\xi^b+\nabla_d\mathcal{G}_{5X}\phi^{cd}\phi_{ca}\phi_{b}\xi^a\xi^b~.
    \end{aligned}
\end{equation}
Although the above expression looks cumbersome, it can be simplified further. 
Thereafter, one can regroup the above expression to get the following\footnote{For a torsionless theory, one has $\phi_{ab}=\phi_{ba}~~, \rm{~and} ~~\phi^a_b=\phi^{~a}_b=\phi^a_{~b}~.$}
\begin{equation}\label{red2}
    \begin{aligned}
        \mathfrak{H}_{\xi\xi}=\mathfrak{H}^{(5)}_{\xi\xi}&=\mathfrak{G}^{(0)}R_{acbd}\phi^c\phi^d\xi^a\xi^b+\mathfrak{G}^{(1)}\phi_{ab}\xi^a\xi^b+\mathfrak{G}^{(2)}\phi_{ca}\phi^{c}_{b}\xi^a\xi^b+\mathfrak{G}^{(3)}_{a}\phi_{b}\xi^a\xi^b\\&~~+\mathfrak{G}^{(4)}\phi^e\phi^d\phi_{db}\phi_{ca}\phi^c_{e}\xi^a\xi^b+\mathfrak{G}^{(5)}\phi^{cd}\phi_{ca}\phi_{db}\xi^a\xi^b+\mathfrak{G}^{(6)}\phi^{c}\phi^{d}\phi_{ca}\phi_{db}\xi^a\xi^b+\mathfrak{F}~,
    \end{aligned}
\end{equation}
with
\begin{equation}
    \begin{aligned}
        \mathfrak{G}^{(0)}&=\mathcal{G}_{4X}-\frac{1}{2}\mathcal{G}_{5X}\Box \phi~,~\mathfrak{G}^{(1)}=\frac{1}{2}\mathcal{G}_{5\phi}\Box \phi~,~\mathfrak{G}^{(2)}=\mathcal{G}_{4X}-\frac{1}{2}\phi^d\nabla_d\mathcal{G}_{5X}-\mathcal{G}_{5X}\Box \phi-\mathcal{G}_{5\phi}~,\\
        \mathfrak{G}_a^{(3)}&=\nabla_{a}\mathcal{G}_3+2\nabla_{c}\mathcal{G}_{4X}\phi^{c}_{a}+2\mathcal{G}_{4X}R_{ca}\phi^c-2\nabla_{a}\mathcal{G}_{4X}\Box \phi+2\mathcal{G}_{4\phi X}\phi^{c}\phi_{ca}+\mathcal{G}_{5X}R_{cd}\phi^{c}\phi^d_{a}\\
        &~~~-\mathcal{G}_{5X}R_{ca}\phi^c\Box \phi+\mathcal{G}_{5X}R_{cdea}\phi^d\phi^{ce}-\nabla^c\mathcal{G}_5R_{ca}-\nabla_d\mathcal{G}_{5X}\Box \phi \phi^{d}_{a}+\nabla_d\mathcal{G}_{5X}\phi^{cd}\phi_{ca}\\
        &~~~+\nabla_{a}\mathcal{G}_{5X}\delta^{ef}_{cd}\phi^c_e\phi^d_f+\frac{1}{2}\nabla_{a}\mathcal{G}_{5\phi}\Box\phi-\nabla_{c}\mathcal{G}_{5\phi}\phi_{a}^{c}-\frac{1}{2}\mathcal{G}_{5\phi X}\phi^c\phi_{ca}\Box \phi+\frac{1}{2}R\nabla_a\mathcal{G}_5~,
    \end{aligned}
\end{equation}
and\footnote[2]{For a smooth function $f(\phi,X)$, we have  $\nabla_c f=f_{\phi}\phi_c-f_X\phi_{ce}\phi^e$.}
\begin{equation}
    \begin{aligned}
    \mathfrak{G}^{(4)}&=-\mathcal{G}_{5XX}~,~
    \mathfrak{G}^{(5)}=~\mathcal{G}_{5X}~,~
    \mathfrak{G}^{(6)}=\mathcal{G}_{5\phi X}+\frac{1}{2}\mathcal{G}_{5XX}\Box \phi-\mathcal{G}_{4XX}~,\\
    \mathfrak{F}&=\nabla^c\mathcal{G}_5\phi^dR_{cabd}\xi^a\xi^b-\nabla_a\mathcal{G}_5R_{bc}\phi^c\xi^a\xi^b+\mathcal{G}_{5X}R_{cdea}\phi^{~d}_b\phi^c\phi^e\xi^a\xi^b~.\\
    \end{aligned}
\end{equation}
Using the fact that $\pounds_{\xi}\phi=0$, and \ref{rel1} we get several trivial terms straightaway such as $\mathfrak{G}^{(1)}\phi_{ab}\xi^a\xi^b$ and $\mathfrak{G}^{(3)}\phi_b\xi^a\xi^b$. The remaining terms in the null-null projection are algebraically cumbersome but straightforward and have been fully analyzed in \ref{appen_A}. By careful algebra, it can therefore be shown that $\mathfrak{H}^{(5)}_{\xi\xi}=0=\mathfrak{H}_{\xi\xi}$. Thus, we have established the vanishing of the null-null projection (for $\pounds_{\xi}\phi=0$) on the Killing horizon, which plays a crucial role for the remaining part of the proof.
\par
This is because, using the equations of motion (\ref{eom}), it now becomes evident that $\mathfrak{T}_{ab}\xi^a\xi^b=0$ on the horizon, implying that $\mathfrak{T}^{ab}\xi_a$ can be expressed as a linear combination of $\xi^b$ and $\mathbf{e}_A^b$ on $\mathcal{H}$. Under the assumption that the minimally coupled matter $\Psi$ satisfies dominant energy condition, we must have $\mathfrak{T}^{ab}\xi_a\propto \xi^b$ implying that $\mathfrak{T}_{ab}\xi^ah^b_e=0$. Hence, for the next part, we shall be looking at the mixed contraction $\mathfrak{H}_{ab}\xi^ah^b_e$ which must identically vanish on $\mathcal{H}$ (and in particular on $\mathscr{C}$), thereby providing us with an identity involving the induced cross-sectional connection $D$, which takes the form of a differential equation for $\kappa$.\par
Before moving forward with the final step of the proof, we shall briefly mention what could potentially happen if we restrict to $\pounds_\xi\phi=\phi_0\neq0~.$ Physically, in that scenario, if we take the adapted coordinate chart where, $\displaystyle{\xi^a=\left(\frac{\partial}{\partial t}\right)^a}$, we get 
\begin{equation}\label{shift}
    \phi(t,y^J)=\phi_0t+C(y^J)~,
\end{equation}
where, $J=1,2,3$.
Such linear-time-dependent scalar field ansatz are often used in Galileon and shift-symmetric scalar-tensor theories\cite{Babichev:2013cya, Minamitsuji:2023nvh}, where they allow the scalar field to break the spacetime symmetry while leaving the metric stationary.\par
In situations like \ref{shift}, the obstruction to $\mathfrak{H}_{\xi\xi}$ being zero arises solely from expressions that previously relied on the assumption $\pounds_\xi\phi=0~$. Without much effort, one can show that 
\begin{equation}\label{null-null_general}
    \mathfrak{H}_{\xi\xi}=\phi_0^2\mathfrak{H}^{(3)}-\kappa\phi_0\mathfrak{H}^{(4)}+\mathfrak{H}^{(5)}_{\xi\xi}~.
\end{equation}
In general, the above expression does not vanish. However, if the general scalar-tensor theory is treated as a regular\footnote[1]{$\mathcal{O}(1/\epsilon)$ order in perturbation is not present.} perturbative deformation around GR w.r.t. the dimensionless perturbation parameter $|\epsilon|\ll1$, i.e.
\begin{equation}\label{perturb}
    \mathcal{G}_{2}=\epsilon \widetilde{\mathcal{G}}_2~,~\mathcal{G}_{3}=\epsilon \widetilde{\mathcal{G}}_3~,~\mathcal{G}_{4}=1+\epsilon \widetilde{\mathcal{G}}_4\quad \rm {and,} \quad\mathcal{G}_5=\epsilon \widetilde{\mathcal{G}}_5~,
\end{equation}
we observe that $\mathfrak{H}^{(3)}, \mathfrak{H}^{(4)}$ and $\mathfrak{H}^{(5)}_{\xi\xi}$ become $\mathcal{O}(\epsilon)$ meaning, $\mathfrak{H}_{\xi\xi}=\mathcal{O}(\epsilon)$. Hence, the perturbative order of $\mathfrak{T}_{\xi\xi}$ is fixed from the above i.e. $\mathcal{O}(\epsilon)$. Hence, we can say that the zeroth-order null-null projection $\mathfrak{T}^{(0)}_{\xi\xi}=0$, which in turn implies that $\mathfrak{T}^{(0)}_{ab}\xi^ah^b_e=0$, i.e. the minimally coupled matter $\Psi$ satisfies DEC at zeroth order. That being said, it is not possible to have $\mathfrak{H}_{\xi\xi}=0$ in generic case when $\pounds_{\xi}\phi\ne\rm const.$
\subsection{Mixed projection: $\mathfrak{H}_{ab}\xi^ah^b_e$}
For the final part of the proof, we shall consider the mixed projection of the equations of motion.
Using \ref{eom}, \ref{group} and \ref{eq18}, we obtain the following
\begin{equation}\label{26}
    \begin{aligned}    \mathfrak{H}_{ab}\xi^ah^b_e=\mathfrak{H}^{(1)}R_{ab}\xi^ah^b_e+\mathfrak{H}^{(5)}\xi^ah^b_e~,
    \end{aligned}
\end{equation}
which can be transformed into a differential equation of $\kappa$, since $D_e\kappa=R_{acbd}\xi^av^ch^{b}_{e}\xi^d=-R_{ab}\xi^ah^{b}_e$~.\\
Next, we contract and group the various contributions according to the number of appearances of $\phi$, $\nabla\phi$ and $\nabla\nabla\phi$. Noting that 
\begin{equation}
    \begin{aligned}
        &\mathfrak{H}^{(5)}_{ab}\xi^ah^b_e\\&=\frac{1}{2}\nabla_{a}\mathcal{G}_3\phi_{b}\xi^ah^b_e+\mathcal{G}_{4X}\phi_{ca}\phi^{c}_{~b}\xi^ah^b_e+\nabla_{c}\mathcal{G}_{4X}\phi^{c}_{~a}\phi_{b}\xi^ah^b_e+\mathcal{G}_{4X}R_{ca}\phi_{b}\phi^c\xi^ah^b_e-\nabla_{a}\mathcal{G}_{4X}\phi_{b}\Box \phi\xi^ah^b_e\\
    &~~+\mathcal{G}_{4X}R_{acbd}\phi^c\phi^d\xi^ah^b_e+\mathcal{G}_{4\phi X}\phi^{c}\phi_{ca}\phi_{b}\xi^ah^b_e-\mathcal{G}_{4XX}\phi^c\phi_{ca}\phi^d\phi_{db}\xi^ah^b_e+\frac{1}{2}\mathcal{G}_{5X}R_{cd}\phi^{c}\phi^d_{~a}\phi_{b}\xi^ah^b_e\\&~~-\frac{1}{2}\mathcal{G}_{5X}R_{ca}\phi_{b}\phi^c\Box \phi\xi^ah^b_e-\frac{1}{2}\mathcal{G}_{5X}R_{acbd}\phi^c\phi^d\Box \phi\xi^ah^b_e+\frac{1}{2}\mathcal{G}_{5X}R_{cdfa}\phi_{b}\phi^d\phi^{cf}\xi^ah^b_e+\mathcal{G}_{5X}R_{cdf(a}\phi^{~d}_{b)}\phi^c\phi^f\xi^ah^b_e\\
    &~~-\frac{1}{2}\nabla_{(a}\left[\mathcal{G}_{5X}\phi^c\right]\phi_{cb)}\Box\phi\xi^ah^b_e+\frac{1}{4}\nabla_{(a}\left[\mathcal{G}_{5\phi}\phi_{b)}\right]\Box\phi\xi^ah^b_e-\nabla_{c}\left[\mathcal{G}_{5\phi}\phi_{(a}\right]\phi_{b)}^{~c}\xi^ah^b_e+\nabla^{c}\mathcal{G}_5\phi^dR_{c(ab)d}\xi^ah^b_e\\&~~-\nabla_{(a}\mathcal{G}_5 G_{b)c}\phi^c\xi^ah^b_e+\frac{1}{2}\nabla_{a}\mathcal{G}_{5X}\phi_{b}\delta^{mn}_{pq}\phi^{p}_{m}\phi^{q}_{n}\xi^ah^b_e-\frac{1}{2}\mathcal{G}_{5X}\Box \phi\phi_{ca}\phi^{c}_{~b}\xi^ah^b_e-\frac{1}{2}\phi^c\nabla_c\mathcal{G}_{5X}\phi_{da}\phi^d_{~b}\xi^ah^b_e\\&~~-\frac{1}{2}\nabla^c\mathcal{G}_5R_{ca}\phi_{b}\xi^ah^b_e+\nabla_c\left[\mathcal{G}_{5X}\phi_d\right]\phi^{c}_{~(a}\phi^{d}_{~b)}\xi^ah^b_e-\frac{1}{2}\nabla_d\mathcal{G}_{5X}\Box \phi \phi^{d}_{~a}\phi_{b}\xi^ah^b_e+\frac{1}{2}\nabla_d\mathcal{G}_{5X}\phi^{cd}\phi_{ca}\phi_{b}\xi^ah^b_e~,
    \end{aligned}
\end{equation}
the above contraction may be regrouped as follows:
\begin{equation}\label{regrp}
    \begin{aligned}
        &\mathfrak{H}^{(5)}_{ab}\xi^ah^b_e\\&=\mathfrak{J}^{(1)}\phi
_{ab}\xi^ah^b_e+\mathfrak{K}^{(1)}\phi_{a}\phi_b\xi^ah^b_e+\mathfrak{K}^{(2)}\phi_{ca}\phi^c_{~b}\xi^ah^b_e+\mathfrak{L}^{(1)}\phi^{c}\phi_{ca}\phi_b\xi^ah^b_e+\mathfrak{L}^{(2)}\phi_{cd}\phi^{c}_{~a}\phi^d_{~b}\xi^ah^b_e\\&~~+\mathfrak{M}^{(1)}\phi_{cd}\phi^d\phi^c_{~a}\phi_b\xi^ah^b_e+\mathfrak{M}^{(2)}\phi^c\phi^d\phi_{ca}\phi_{db}\xi^ah^b_e+\mathfrak{N}^{(1)}\phi^{cf}\phi_f\phi^d\phi_{ca}\phi_{db}\xi^ah^b_e+\mathfrak{N}^{(2)}\phi^{f}\phi_{fd}\phi^{cd}\phi_{ca}\phi_b\xi^ah^b_e\\&~~+\mathfrak{N}^{(3)}\phi_{cf}\phi^c\phi^f\phi_{da}\phi^d_{~b}\xi^ah^b_e+\mathfrak{N}^{(4)}\phi^{cf}\phi_f\phi^d\phi_{da}\phi_{cb}\xi^ah^b_e+\mathfrak{R}_e\equiv\mathfrak{Z}_e+\mathfrak{R}_e~.
    \end{aligned}
\end{equation}
In the above expression, $\mathfrak{Z}_e$ forms the trivial part of the projection that identically vanishes (for details, cf. \ref{B}) under the assumption that $\pounds_{\xi}\phi=0$, with
\begin{equation}
    \begin{aligned}
        \mathfrak{J}^{(1)}&=\frac{1}{4}\mathcal{G}_{5\phi}\Box \phi~;~
        \mathfrak{K}^{(1)}=\frac{1}{2}\mathcal{G}_{3\phi}-\mathcal{G}_{4X\phi}\Box \phi+\frac{1}{2}\mathcal{G}_{5X\phi}\delta^{mn}_{pq}\phi^{p}_{m}\phi^{q}_{n}+\frac{1}{4}\mathcal{G}_{5\phi\phi}\Box\phi~;\\
        \mathfrak{K}^{(2)}&=\mathcal{G}_{4X}-\mathcal{G}_{5X}\Box \phi-\frac{1}{2}\mathcal{G}_{5X\phi}\phi^i\phi_i-\mathcal{G}_{5\phi}~;\\
        \mathfrak{L}^{(1)}&=-\frac{1}{2}\mathcal{G}_{3X}+2\mathcal{G}_{4\phi X}+\mathcal{G}_{4XX}\Box \phi-\frac{1}{2}\mathcal{G}_{5XX}\delta^{mn}_{pq}\phi^{p}_{m}\phi^{q}_{n}-\frac{3}{8}\mathcal{G}_{5\phi X}\Box \phi-\frac{1}{2}\mathcal{G}_{5\phi\phi}~;\\
        \mathfrak{L}^{(2)}&=\mathcal{G}_{5X}~;~
        \mathfrak{M}^{(1)}=\mathfrak{M}^{(2)}=-\mathcal{G}_{4XX}+\frac{1}{2}\mathcal{G}_{5XX}\Box \phi+\mathcal{G}_{5X\phi}~;\\
        \mathfrak{N}^{(1)}&=\mathfrak{N}^{(2)}=\mathfrak{N}^{(4)}=-\frac{1}{2}\mathcal{G}_{5XX}~;~
        \mathfrak{N}^{(3)}=\frac{1}{2}\mathcal{G}_{5XX}~.\\
    \end{aligned}
\end{equation}
The quantity $\mathfrak{R}_e$ above captures the curvature-dependent part of the mixed projection $\mathfrak{H}^{(5)}\xi^ah^b_e$, which eventually gives rise to the non-vanishing contributions. It is defined as
\begin{equation}\label{curv}
\begin{aligned}
            \mathfrak{R}_e&=\left(\mathcal{G}_{4X}-\frac{1}{2}\mathcal{G}_{5X}\Box \phi\right)R_{acbd}\phi^c\phi^d\xi^ah^b_e+\left(\mathcal{G}_{4X}-\frac{1}{2}\mathcal{G}_{5X}\Box \phi\right)R_{ca}\phi_{b}\phi^c\xi^ah^b_e+\frac{1}{2}\mathcal{G}_{5X}R_{cd}\phi^{c}\phi^d_{~a}\phi_{b}\xi^ah^b_e\\&+\frac{1}{2}\mathcal{G}_{5X}R_{cdfa}\phi_{b}\phi^d\phi^{cf}\xi^ah^b_e+\mathcal{G}_{5X}R_{cdf(a}\phi^{~d}_{b)}\phi^c\phi^f\xi^ah^b_e+\nabla^{c}\mathcal{G}_5\phi^dR_{c(ab)d}\xi^ah^b_e-\nabla_{(a}\mathcal{G}_5 G_{b)c}\phi^c\xi^ah^b_e\\&-\frac{1}{2}\nabla^c\mathcal{G}_5R_{ca}\phi_{b}\xi^ah^b_e~.
\end{aligned}
\end{equation}
Even in the above curvature-dependent part, certain terms vanish due to geometric identities on $\mathcal{H}$ \ref{geom_identities}, along with the Killing-symmetry assumption for the scalar field. Once all such cancellations are taken into account, we are left with the following equation
\begin{equation}\label{31}
    \mathfrak{H}_{ab}\xi^ah^b_e=-\mathfrak{H}^{(1)}D_e\kappa+\mathfrak{R}_e~
\end{equation}
where the rhs of \ref{31} can now be set to zero using the dominant energy condition assumed for minimally coupled matter fields and the equations of motion.
\subsubsection{Obtaining the equation for $D_e \kappa$}
After collecting the non-zero terms in $\mathfrak{R}_e$ (as worked out in \ref{C}), we obtain the simplified version of the equation for $D_e \kappa$, ({when $\phi$ is Lie dragged}) which is
\begin{equation}\label{fineq}
\begin{aligned}
    &\mathfrak{H}_{ab}\xi^ah^b_e\\&=-\mathfrak{H}^{(1)}D_e\kappa-(\mathcal{G}_{4X}-\frac{1}{2}\mathcal{G}_{5X}\Box\phi-\mathcal{G}_{5X}\Phi-\mathcal{G}_{5\phi})(D^m\phi)(D_m\kappa)(D_e\phi)-\frac{1}{2}\mathcal{G}_{5X}(D_m\kappa)(D_n\phi)(D_e\phi)(D^m D^n \phi)\\
    &~-\frac{1}{2}\mathcal{G}_{5X}(D^m\phi)(D_m\kappa)(D_eD_n\phi)(D^n\phi)=0
\end{aligned}
\end{equation}
with $\Phi\equiv \phi_{ab}v^a\xi^b$.\par
We can now use the above expression to examine a certain subclass of generic scalar-tensor theories for which $\mathcal{G}_5=\mathcal{G}_5(\phi)$ and $\mathcal{G}_4=Xh(\phi)$, where $h(\phi)$ is a differentiable function such that $\mathcal{G}_{5\phi}=h(\phi)$. In such cases, $D_e\kappa$ may or may not be zero, so one cannot simply infer the zeroth law from equation \ref{fineq}. This is due to the fact that $\mathfrak{H}^{(1)}=Xh(\phi)+\frac{1}{2}\phi^c\phi_ch(\phi)$ identically vanishes irrespective of $h(\phi)$.
However, an interesting observation arises when we consider the consequences of GW constraining Horndeski theories in cosmological backgrounds\cite{Baker:2017hug, DeFelice:2011bh}, following the multimessenger detection of GW170817 \cite{Margutti:2020xbo}, together with its electromagnetic counterpart GRB170817A \cite{Kreisch:2017uet}.
Here, it is to be mentioned that the effective Planck mass is given in terms of the Horndeski coupling functions as $m_*=2(\mathcal{G}_4-2X\mathcal{G}_{4X}+X\mathcal{G}_{5\phi}-\dot{\phi}HX\mathcal{G}_{5X})$~, and the deviation $\alpha_{\rm T}$ related to the speed of GW, which is given as $c_{\rm T}^2=1+\alpha_{\rm T}$~ becomes
\begin{equation}
    m_*^2\alpha_{\rm T}=2X\left(2\mathcal{G}_{4X}-2\mathcal{G}_{5\phi}-(\ddot{\phi}-\dot{\phi}H)\mathcal{G}_{5X}\right)~.
\end{equation}
The demand $\alpha_{\rm T}\sim 0$ can be fulfilled by assuming $\mathcal{G}_{5}=0$ and $\mathcal{G}_{4X}=0$. This naturally constrains this class of theories and ensures the luminal speed of GW, avoiding non-trivial cancellations. Now, in such cases, we observe that \ref{fineq} reduces to 
\begin{equation}
    \mathfrak{H}^{(1)}D_e\kappa=0~,
\end{equation}
thereby implying $D_e\kappa=0$ as $\mathfrak{H}^{(1)}=\mathcal{G}_4\ne0$. Hence, in this case, we recover the zeroth law, without needing a perturbative extension around GR.\par Finally, we study the Horndeski theories in their most general form, without imposing any restriction on the Horndeski functions. We stress, however, that our proof is based on a perturbative expansion around GR and therefore requires a well-defined perturbative connection to the GR limit. The present approach therefore fails for theories which do not allow such a perturbative expansion.
As mentioned earlier \ref{perturb}, we consider the general Horndeski functions as follows
\begin{equation*}
    \mathcal{G}_i=\mathcal{G}_{i,0}+\epsilon\delta\mathcal{G}_i~,
\end{equation*}
where $\delta\mathcal{G}_i=\widetilde{\mathcal{G}}_i$ for \(i=2,3,4,5\), with $\mathcal{G}_{4,0}=1$, while $\mathcal{G}_{i,0}=0$ for \(i\neq4\).
Thereafter, one can rewrite the equation \ref{fineq} as a matrix equation
\begin{equation}\label{zero}
    \mathcal{X}_e=\epsilon \sum_m\mathcal{B}^m_e \mathcal{X}_m~,
\end{equation}
where, we have defined $\mathcal{X}_e= D_e \kappa$ and 
\begin{equation}
    \mathcal{B}^m_e = \left[-\widetilde{\mathcal{G}}
_{4}-\frac{1}{2}\nabla_c\widetilde{\mathcal{G}}_5\phi^c\right]\delta^m_e+\left[\frac{1}{2}\widetilde{\mathcal{G}}_{5X}\Box \phi-\widetilde{\mathcal{G}}_{4X}+\widetilde{\mathcal{G}}_{5\phi}+\widetilde{\mathcal{G}}_{5X}\Phi\right](D^m\phi)(D_e \phi)-\frac{1}{2}\widetilde{\mathcal{G}}_{5X}(D_a \phi)(D^a(D^m \phi D_e\phi))~,
\end{equation}
as the matrix elements.
Following the argument in Ref. \cite{Ghosh:2020dkk}, we consider perturbative expansions\footnote[1]{Regime of validity for such expansion in general would be $|\epsilon^{k+1}\mathcal{X}_e^{(k+1)}|\ll|\epsilon^{k}\mathcal{X}_e^{(k)}|$, and $|\epsilon^{k+1}\mathcal{B}_e^{m,(k+1)}|\ll|\epsilon^{k}\mathcal{B}_e^{m, (k)}|,~\forall m$, for $k\ge0$. A sufficient, though not necessary, assumption is $|\epsilon|\ll1$, which provides controlled corrections to the leading order.} in powers of $\epsilon$.
\begin{equation}\label{zeros}
    \mathcal{X}_e=\sum_{k\ge0}\mathcal{X}_e^{(k)}\epsilon^k~\quad ,\quad \mathcal{B}^m_e=\sum_{k\ge0}\mathcal{B}_e^{m,(k)}\epsilon^k~,
\end{equation}
which after plugging in \ref{zero}, provides the following recursion relation 
\begin{equation}
    \mathcal{X}^{(k)}_e=\sum_m\sum_{l=0}^{k-1}\mathcal{B}^{m,(l)}_e\mathcal{X}^{(k-l-1)}_m~\quad\rm{and}\quad k\ge1~,
\end{equation}
with $\mathcal{X}^{(0)}_e=0$ (as the perturbative order of $\mathcal{X}_e$ is fixed to be $\mathcal{O}(\epsilon)$ from \ref{zero}). The above system of recursion relations can be solved by induction \cite{Sang:2021rla} to obtain $\mathcal{X}^{(k)}_e=0$ for all $k\in \mathbb{N}^+$, implying that $\mathcal{X}_e=D_e\kappa=0$, thereby concluding our proof.\par
We may now ask: \textit{To what extent is the zeroth law valid in less restrictive cases?} To answer this, we start from the assumption: $\pounds_{\xi}\phi=\phi_0\ne0$ (\textit{restrictive case}). The preceding analysis in \ref{null_null_proj} established the fact that only the zeroth-order mixed projection vanishes i.e., $\mathfrak{T}^{(0)}_{ab}\xi^ah^b_e=0$ (for such restricted case), with the result obtained within a perturbative treatment. Quite similar to \ref{31}, it is straightforward to obtain the following equation
\begin{equation}
    \frac{1}{2}\mathfrak{T}_{ab}\xi^ah^b_e=-\mathfrak{H}^{(1)}D_e\kappa+\phi_0\mathfrak{H}^{(3)}(D_e\phi)+\mathfrak{H}^{(4)}\phi_{ab}\xi^ah^{b}_e+\mathfrak{Z}_e+\mathfrak{R}_e~.
\end{equation}
$\mathfrak{Z}_e$ now becomes non-trivial as it contains all the non-zero contributions from the null-transverse projection which were previously set to zero due to $\pounds_{\xi}\phi=0$. It is worth recalling that $\mathfrak{H}^{(3)},\mathfrak{H}^{(4)},\mathfrak{Z}_e,\mathfrak{R}_e$ are all $\mathcal{O}(\epsilon)$. Hence, it is only fitting to claim that under such restrictions on the scalar field $\phi$, 
\begin{equation}(\mathfrak{H}^{(1)}D_e\kappa)^{(0)}=0~\quad\text{(equating the coefficients of $\epsilon^0$)}~,
\end{equation}
which subsequently implies that $(D_e\kappa)^{(0)}=0$. Here, the surface gravity can be expanded in terms of $\epsilon$ as $\displaystyle{\kappa}=\sum_{k\ge0}\kappa^{(k)}\epsilon^k$, therefore allowing us to identify the above result as $D_e\kappa^{(0)}=0$, which we shall refer to as the \textit{zeroth-order zeroth law}. Thus, for such restrictive cases, only the zeroth-order contribution to the surface gravity remains constant within the perturbative analysis. The higher orders in perturbation, namely $\kappa^{(k)}$ for $k\ge1$ may not be constant. In generic cases, beyond the above-mentioned restrictions, the constancy of surface gravity is no longer guaranteed.\par

Before concluding, we emphasize that the condition $\pounds_\xi\phi=0$ is not an arbitrary technical restriction but the natural statement of stationarity for the full gravitational-scalar system for the following reasons.
First, a stationary black hole solution of a theory with fields $(g_{ab},\phi,\Psi)$ is, by definition, one for which the Killing field $\xi^a$ generates a symmetry of the \textit{entire} configuration, $\pounds_\xi g_{ab}=\pounds_\xi\phi=\pounds_\xi\Psi=0$. This is exactly the condition already implicitly imposed on the minimally coupled matter $\Psi$ whenever the dominant energy condition is invoked on $\mathfrak{T}_{ab}$, and there is no principled reason to treat the scalar sector differently from the matter sector (cf.\ the treatment of the scalar field in Hawking's analysis of stationary black holes in Brans--Dicke theory \cite{Hawking:1972qk}).
Second, the condition is essentially forced by the field equations whenever the couplings depend explicitly on $\phi$. If $g_{ab}$ and $\mathfrak{T}_{ab}$ are stationary and $\pounds_\xi\phi=\phi_0$ is constant, then $\pounds_\xi\phi_a=\nabla_a\phi_0=0$ and $\pounds_\xi X=0$, so the only way $\mathfrak{H}_{ab}$ can change along $\xi^a$ is through the explicit dependence of the $\mathcal{G}_i(\phi,X)$ on $\phi$. Applying $\pounds_\xi$ to \ref{eom} then gives $\phi_0\,\partial_\phi\mathfrak{H}_{ab}\big|_{\rm explicit}=0$, which for a generic theory with $\mathcal{G}_{i\phi}\neq0$ can only be satisfied by $\phi_0=0$. A non-vanishing $\phi_0$ is therefore compatible with a stationary metric only for shift-symmetric theories, which is precisely the setting of Case B below.
Third, for an isolated black hole with the standard asymptotic boundary condition $\phi\to\phi_\infty=\text{const}$, one has $\xi^a\phi_a\to0$ at infinity; since $\pounds_\xi\phi=\phi_0$ is constant, this forces $\phi_0=0$ everywhere. Solutions with $\phi_0\neq0$ require a scalar that grows linearly in time even far from the hole, i.e.\ cosmological-type boundary conditions \cite{Babichev:2013cya}.
Thus the Lie-dragged condition holds for the generic stationary black hole, and it is under this condition that the zeroth law is established in Case A, subject to the dominant energy condition and the perturbative (or luminal-GW) conditions discussed in \ref{sec:zeroth_law}. Relaxing it requires both a shift-symmetric theory and non-standard asymptotics, and the physics changes in two ways. When the scalar is Lie-dragged, it is static along every horizon generator, so nothing scalar flows through the horizon and the null-null projection of the field equations vanishes identically. When the scalar instead grows linearly along the Killing time, it carries a steady kinetic flux across the horizon. The field equations then no longer balance on their own, and the balance can be restored only if this flux is small, which is why a perturbative expansion about GR becomes essential. The same flux means that the matter cannot be required to satisfy the dominant energy condition exactly. It can be required to do so only at the leading order, where the scalar's influence has not yet entered. Moreover, the scalar's time dependence contributes additional terms to the transverse projection of the field equations, terms that were absent for a Lie-dragged scalar. These additional terms enter only at higher orders in the deformation away from GR. The projection therefore fixes the leading-order surface gravity, so the horizon has a well-defined zeroth-order temperature, but it cannot guarantee that corrections to this temperature are uniform over the horizon. If the scalar's time dependence is more general still, the horizon is no longer in equilibrium with its scalar hair, the flux is uncontrolled, and the argument cannot begin even perturbatively. It is therefore the equilibrium between horizon and scalar, not the complexity of the coupling, that decides how much of the zeroth law survives.

\section{Conclusion}\label{sec:conclusion}
In this paper, we prove the zeroth law for (full/general) Horndeski theories expressed in terms of the DGSZ action. The central physical question motivating this analysis is a simple one: does a stationary black hole in a theory with an arbitrary non-minimal scalar-curvature coupling still possess a single, well-defined horizon temperature, exactly as required for the zeroth law of ordinary thermodynamics to carry over meaningfully to the gravitational setting? Our analysis shows that the answer depends sensitively on how the scalar field $\phi$ itself behaves relative to the horizon's own symmetry - a dependence that has no analog in GR, where no such additional field is present to begin with. Based on our findings, we shall classify our results into three distinct regimes of validity, organized by the behavior of $\pounds_\xi\phi$:\par
{\textit{Case A (Lie dragged scalar)}}: When $\pounds_{\xi}\phi=0$, the basic setup for the proof assumes that the non-minimally coupled field $\phi$, has the same symmetries as the metric. We further found that the surface gravity $\kappa$, must satisfy an equation of the form $-\mathfrak{H}^{(1)}D_e\kappa+\mathfrak{R}_e=0$, provided the DEC holds on minimally coupled matter. The vanishing of the two-covariant derivative of $\kappa$ is equivalent to the invariance of surface gravity on the horizon cross-section $\mathscr{C}$.  When $\mathcal{G}_{5\phi}=h(\phi)$, and $\mathcal{G}_4(\phi,X)=Xh(\phi)$, the final mixed projection does not suffice to establish the zeroth law. In general, while deriving the final expression, one needs to realize the scalar-tensor theory as a perturbative extension assuming a smooth limit to GR as ($\epsilon \to 0$). However, this assumption need not be strictly imposed in all cases. In particular, the observational constraint that gravitational waves propagate at luminal speed\cite{Baker:2017hug} can, in certain special scenarios discussed above ($\mathcal{G}_5=\mathcal{G}_{4X}=0$), effectively render the perturbative procedure unnecessary.\par

{\textit{Case B (Shift-symmetric case)}}: In situations where \(\pounds_{\xi}\phi=\phi_0\ne0\), (corresponding to linearly time-dependent scalar hair characteristic of known stealth\footnote[1]{For a stealth solution, the metric is precisely its GR background value, so there are no higher-order metric corrections and therefore $\kappa=\kappa^{(0)}$. Thus the zeroth law holds geometrically and is consistent with $D_e\kappa^{(0)}=0$.} and hairy black hole solutions in shift-symmetric and Galileon theories \cite{Babichev:2013cya,Kobayashi:2014eva,Babichev:2016rlq}) a perturbative treatment is required. Such solutions evade the usual no-hair expectations precisely because the scalar field does not inherit the stationary symmetry of the horizon, and therefore constitute a natural setting in which the validity of the zeroth law may be expected to be nontrivial. In this case, however, the final mixed projection is not sufficient to establish the zeroth law at all orders in the perturbative expansion. Rather, as expected from the perturbative structure, it guarantees only the constancy of the leading-order contribution to the surface gravity, \(\kappa^{(0)}\). We regard this \textit{zeroth-order zeroth law} as a genuinely non-trivial result in its own right: even when the scalar actively breaks the symmetry that the standard proof relies on, surface gravity still retains a well-defined leading-order value, and the possible non-constancy of sub-leading orders is confined strictly to higher orders in the deformation away from GR.\par

{\textit{Case C (Generic case)}}: Lastly, when $\pounds_{\xi}\phi\ne0$, the null-null projection $\mathfrak{H}_{\xi\xi}$ does not vanish, and the zeroth law cannot be proven even after a perturbative treatment. This is not merely a technical gap in our method. Rather, it reflects the fact that an arbitrarily time-dependent scalar profile along the horizon generators provides a genuine, uncontrolled flux onto the horizon through the $\phi_a\phi_b\xi^a\xi^b$ and $\phi_{ab}\xi^a\xi^b$ terms. So the breakdown of the proof plausibly tracks a genuine physical obstruction rather than only a limitation of the perturbative expansion.\par
\par
Our results therefore clarify the conditions under which the zeroth law can be regarded as universal and the extent to which it is valid, thereby extending it to a broader class of scalar-tensor theories. The three cases stated above trace a single underlying theme: it is specifically the relationship between the scalar hair and the horizon's own stationary symmetry, not the complexity of the non-minimal coupling $\mathcal{G}_i(\phi,X)$ by itself, that governs whether the zeroth law survives. A theory with an arbitrarily intricate $\mathcal{G}_5$ sector still obeys the zeroth law exactly, with no perturbative caveat at all, provided $\phi$ respects the horizon symmetry. However, when the luminal-speed constraint holds, perturbative expansion is not necessary. Conversely, even the mildest relaxation of that symmetry, to a mere constant $\pounds_\xi\phi=\phi_0$, is enough to confine the result to leading order in a deformation parameter. These results contribute to the existing discussion on the universality of the black hole zeroth law. In particular, a recent work\cite{Ghosh:2026ixw} has pointed out that the mechanism responsible for keeping the surface gravity constant might differ from one gravitational theory to another. Our study offers another perspective within the scalar-tensor framework, emphasizing the effect of the scalar-field configuration and the associated projections of the field equations, and suggests that, at least within general Horndeski gravity, the symmetry properties of the matter content sourcing the non-minimal coupling, may be as decisive for black hole thermodynamics as the structure of the gravitational action itself.

\section*{Acknowledgement}
We gratefully acknowledge Sumanta Chakraborty for his valuable comments on this work.
Research of RP is funded by the University Grants Commission (UGC), Government of India (Ref: 211610000414). Research of KB is supported by the New Faculty Seed Grant of BITS Pilani Dubai Campus.
\section*{Appendix}
\appendix
\labelformat{section}{Appendix #1} 
\labelformat{subsection}{Appendix #1}
\section{Vanishing of null-null projection}\label{appen_A}
In this appendix, we provide the details to the calculations that lead to the vanishing of the null-null projection of the gravitational field equations on the Killing horizon $\mathcal{H}$, i.e. $\mathfrak{H}_{\xi\xi}=0$. Apart from the obvious trivial expressions mentioned in the main text, we explicitly calculate several other terms for completeness.\par
To begin with, we handle those terms in \ref{red2} that are proportional to $R_{acbd}\xi^a\xi^b\phi^c\phi^d$. On the horizon $\mathcal{H}$, we have the inverse metric given as
\begin{equation}\label{dec}
    g^{ab}=-2\xi^{(a}v^{b)}+h^{ab}~.
\end{equation}
Thus, some simplifications can be immediately made for
\begin{equation}\label{16}
\begin{aligned}
        R_{acbd}\xi^a\xi^b\phi^c\phi^d&= R_{acbd}\xi^a\xi^b\left(-\xi^cv_e-\xi_ev^c+h^c_e\right)\left(-\xi^dv_f-\xi_fv^d+h^d_f\right)\phi^e\phi^f\\
        &=R_{acbd}\xi^a\xi^bh^c_eh^d_f\phi^e \phi^f=(R_{acbd}\xi^a\xi^b\mathbf{e}^{c}_A\mathbf{e}^{d}_C)\phi_e \phi_f\mathbf{e}^{e}_B\mathbf{e}^{f}_D\delta^{AB}\delta^{CD}
\end{aligned}
\end{equation}
where, we have used the symmetry properties of $R_{acbd}$ and the fact that $\pounds_{\xi}\phi=0$. Further, because the expansion and shear vanish on the horizon, we have $R_{acbd}\xi^a\xi^b\mathbf{e}^{c}_A\mathbf{e}^{d}_C=0$. Thus, we have vanishing contributions from the term $\mathfrak{G}^{(0)}R_{acbd}\xi^a\xi^b\phi^c\phi^d$ in \ref{red2}. Similarly, for the third term in \ref{red2}, we have 
\begin{equation}
    \begin{aligned}
        \xi^a\xi^b\phi_{ca}\phi^c_{~b}&=\xi^a\xi^bg^{cd}\phi_{ca}\phi_{db}\\
        &=\xi^a\xi^bh^{cd}\phi_{ca}\phi_{db}-2\xi^a\xi^b\xi^cv^d\phi_{ca}\phi_{db}\\
        &=\xi^a\xi^bh^{cd}\phi_{ca}\phi_{db}~,
    \end{aligned}
\end{equation}
where in arriving at the last step, we use the relation \ref{rel1}. For the remaining term above, one can look at the following expansion
\begin{equation}\label{eq18}
    \xi^bh^d_c\phi_{db}=-(h^d_c\nabla_d \xi^b)\phi_b
\end{equation}
as $\pounds_{\xi}\phi=0$. Then, immediately one can simplify further since
\begin{equation}\label{eq19}
    \begin{aligned}
        (h^d_c\nabla_d \xi^b)\phi_b&=-(\delta^d_c+\xi^dv_c+\xi_cv^d)\nabla_d\xi^b\phi_b\\
        &=-\phi^b(\nabla_c\xi_b+\kappa v_c\xi_b+v^d\xi_c\nabla_d\xi_b)\\
        &=-\phi^b(\nabla_c\xi_b+\kappa v_c\xi_b+v^d\xi_d\nabla_c\xi_b+v^d\xi_b\nabla_d\xi_c)\\
        &=-\phi^b(\nabla_c\xi_b-\nabla_c\xi_b+\xi_b(\kappa v_c+v^d\nabla_d\xi_c))=0~.
    \end{aligned}
\end{equation}
Here, in the second-to-last line, we used the fact that the generator of the Killing horizon orthogonal to the hypersurface must satisfy the irrotationality condition\footnote[3]{Frobenius' theorem}, i.e. $\xi_{[b}\nabla_d\xi_{c]}=0$. As a result, in \ref{red2}, the term proportional to $\mathfrak{G}^{(2)}$ vanish. Similarly, we can examine the fate of the terms proportional to $\mathfrak{G}^{(4)}$, $\mathfrak{G}^{(5)}$, and $\mathfrak{G}^{(6)}$, respectively.\\ Firstly, we shall expand the following
\begin{equation}\label{no1}
    \begin{aligned}  \xi^a\xi^b\phi_e\phi^d\phi^{ce}\phi_{ca}\phi_{db}&=\xi^a\xi^b\phi_e\phi_f\phi_{mn}\phi_{ca}\phi_{db}g^{cm}g^{ne}g^{df}=\xi^a\xi^b\phi_e\phi_f\phi_{mn}\phi_{ca}\phi_{db}(-\xi^dv^f-\xi^fv^d+h^{df})g^{ne}g^{cm}\\
    &=\xi^a\left(\xi^bh^d_f\phi_{db}\right)\phi_e\phi^f\phi_{mn}g^{ne}g^{cm}=0~.
    \end{aligned}
\end{equation}
Next, we shall look at the expansion of $\phi^{cd}\phi_{ca}\phi_{db}\xi^a\xi^b$ which is as follows:
\begin{equation}\label{no2}
\begin{aligned} \phi_{cd}\phi^{c}_{~a}\phi^{d}_{~b}\xi^a\xi^b&=\xi^a\xi^b\phi_{cd}\phi_{ea}\phi_{fb}g^{ce}g^{df}\\
&=\xi^a\xi^b\phi_{cd}\phi_{ea}\phi_{fb}(-\xi^cv^e-\xi^ev^c+h^{ce})(-\xi^dv^f-\xi^fv^d+h^{df})\\
&=\xi^a\xi^bh^{ce}h^f_d\phi_{ea}\phi_{fb}\phi_{c}^{~d}-\xi^a\xi^b\xi^dv^fh^c_e\phi^{e}_{~a}\phi_{fb}\phi_{cd}-\xi^a\xi^b\xi^cv^eh^d_f\phi_{ea}\phi^f_{~b}\phi_{cd}=0~.
\end{aligned}
\end{equation}
Lastly, we look at the expansion for $\phi^{c}\phi^d\phi_{ca}\phi_{db}\xi^a\xi^b$ which yields:
\begin{equation}\label{no3}
    \begin{aligned}  \phi^{c}\phi^d\phi_{ca}\phi_{db}\xi^a\xi^b&=\xi^a\xi^b\phi_e\phi_f\phi_{ca}\phi_{db}g^{ce}g^{df}\\
    &=\xi^a\xi^b\phi_e\phi_f\phi_{ca}\phi_{db}(-\xi^cv^e-\xi^ev^c+h^{ce})(-\xi^dv^f-\xi^fv^d+h^{df})\\
    &=\xi^a\xi^bh^{c}_eh^{df}\phi^e\phi_f\phi_{ca}\phi_{db}=0~.\\
    \end{aligned}
\end{equation}
In obtaining \ref{no1}, \ref{no2}, and \ref{no3}, we have used the previously obtained identities $ \xi^bh^d_c\phi_{db}=0$, and, $\xi^a\xi^b\phi_{ab}=0$ and the fact that the metric decomposition as \ref{dec} is possible on the horizon $\mathcal{H}$. Thus, we find that the terms with $\mathfrak{G}^{(4)}$, $\mathfrak{G}^{(5)}$, and $\mathfrak{G}^{(6)}$ vanish identically.
\par
 Eventually, the contraction $\mathfrak{H}_{ab}\xi^a\xi^b$ only leads to terms in $\mathfrak{F}$ which are all dependent on the curvature tensor. For the analysis of $\mathfrak{F}$, we begin with the expansion for the quantity $\nabla_a\mathcal{G}_5R_{bc}\phi^c\xi^a\xi^b$, which is as follows
 \begin{equation}
     \begin{aligned}
         \nabla_a\mathcal{G}_5R_{bc}\phi^c\xi^a\xi^b&=-\mathcal{G}_{5X}\phi_{ad}\phi^dR_{bc}\phi^c\xi^a\xi^b~.\\
     \end{aligned}
 \end{equation}
 Then, further simplification can be done as
 \begin{equation}
     \begin{aligned}
         \phi_{ad}\phi^dR_{bc}\phi^c\xi^a\xi^b&=\xi^a\xi^b R_{bc}\phi_{ad}\phi_e\phi_fg^{de}g^{cf}\\
         &=\xi^a\xi^b R_{bc}\phi_{ad}\phi_e\phi_f(-\xi^dv^e-\xi^ev^d+h^{de})(-\xi^cv^f-\xi^fv^c+h^{cf})\\
         &=\xi^a\xi^b R_{bc}\phi_{ad}\phi_e\phi_fh^{de}(-\xi^cv^f-\xi^fv^c+h^{cf})\\
         &=\xi^a\xi^b R_{bc}\phi_{ad}\phi_e\phi^fh^{de}h^c_f=0~.
     \end{aligned}
 \end{equation}
 For the above simplification, we have used the relation $R_{\xi\xi}=0$ on $\mathcal{H}$, and the previously mentioned identities. The term $\mathcal{G}_{5X}R_{cdea}\phi^{~d}_b\phi^c\phi^e\xi^a\xi^b$ in $\mathfrak{F}$ indeed vanishes, as one can look at the following expansion:
 \begin{equation}
     \begin{aligned}
         &R_{cdea}\phi^{~d}_b\phi^c\phi^e\xi^a\xi^b=R_{cdea}\phi_{bf}\phi_m\phi_n\xi^a\xi^bg^{df}g^{cm}g^{en}\\
         &=-R_{cdea}\xi^a\xi^b\phi^m\phi^n\xi^dv^fh^c_mh^e_n\phi_{bf}+R^{d}_{~cea}\xi^a\xi^b\phi^m\phi^n\xi^cv_mh^{f}_dh^e_n\phi_{bf}+R_{cdea}\xi^a\xi^b\phi^m\phi^nh^{fd}h^c_mh^e_n\phi_{bf}\\
         &=-R_{cdea}\xi^a\xi^b\phi_m\phi_n\xi^dv^fh^{cm}h^{en}\phi_{bf}+R_{cdea}\xi^a\xi^b\phi_m\phi_nh^{fd}h^{cm}h^{en}\phi_{bf}\\
         &=-(R_{cdea}\xi^d\xi^a\mathbf{e}^c_A\mathbf{e}^e_C)\xi^b\phi_m\phi_nv^f\phi_{bf}\mathbf{e}^m_B\mathbf{e}^n_D\delta^{AB}\delta^{CD}-(R_{aecd}\xi^a\mathbf{e}^e_A\mathbf{e}^c_C\mathbf{e}^d_E)\xi^b\phi_m\phi_n\phi_{bf}\mathbf{e}^n_B\mathbf{e}^m_D\mathbf{e}^f_F\delta^{AB}\delta^{CD}\delta^{EF}\\
         &=0~.
     \end{aligned}
 \end{equation}
Besides the previously used identities, here we use the relations: \\$R_{cdea}\xi^d\xi^a\mathbf{e}^c_A\mathbf{e}^e_C=0$, and $R_{aecd}\xi^b\mathbf{e}^e_A\mathbf{e}^c_C\mathbf{e}^d_E=0$ on the horizon $\mathcal{H}$.
Finally, as $\nabla^c\mathcal{G}_5\phi^dR_{cabd}\xi^a\xi^b=0$, for more details, see the next subsection i.e. \ref{A}, $\mathfrak{F}$ identically vanishes and we obtain $\mathfrak{H}_{\xi\xi}=0$.
\subsection{Expansion of $\nabla^c\mathcal{G}_5\phi^d R_{cabd}\xi^a\xi^b$ in $\mathfrak{F}$}\label{A}
Although less obvious, the expansion of $\nabla^c\mathcal{G}_5\phi^d R_{cabd}\xi^a\xi^b$ vanishes.
For a given differentiable function $f=f(\phi,X)$, one has
\begin{equation}
    \nabla^c f=f_{\phi}\nabla^c \phi +f_{X}\nabla^c X=f_{\phi} \phi^c -f_{X}\phi^{ce}\phi_e~.
\end{equation}
Thus, $\nabla^c\mathcal{G}_5=\mathcal{G}_{5\phi}\phi^c-\mathcal{G}_{5X}\phi^{ce}\phi_e$, which yields
\begin{equation}\label{28}
    \nabla^c\mathcal{G}_5\phi^d R_{cabd}\xi^a\xi^b=\mathcal{G}_{5\phi}\phi^c\phi^d R_{cabd}\xi^a\xi^b-\mathcal{G}_{5X}\phi^{ce}\phi_e\phi^d R_{cabd}\xi^a\xi^b~.
\end{equation}
Since, for the first term in the above expression
\begin{equation}
    \begin{aligned}
        \phi^c\phi^d R_{cabd}\xi^a\xi^b=-\phi^c\phi^d R_{cbda}\xi^a\xi^b-\phi^c\phi^d R_{cdab}\xi^a\xi^b=0~,
    \end{aligned}
\end{equation}
from the fact that $R_{cabd}+ R_{cbda}+R_{cdab}=0$ and \ref{16}, we are only left with the second term.
Again, we may have 
\begin{equation}
    \begin{aligned}
        &\phi^{ce}\phi_e\phi^d R_{cabd}\xi^a\xi^b=-\phi^{ce}\phi_e\phi^d R_{cbda}\xi^a\xi^b=-g^{cf}g^{he}\phi_{fh}\phi_e\phi_m R_{cbda}\xi^a\xi^bg^{md}\\
        &=-(-\xi^cv^f-\xi^fv^c+h^{cf})(-\xi^hv^e-\xi^ev^h+h^{he})\phi_{fh}\phi_e\phi_m R_{cbda}\xi^a\xi^bg^{md}\\
        &=(\xi^fv^c-h^{cf})(\xi^hv^e-h^{he})\phi_{fh}\phi_e\phi_m R_{cbda}\xi^a\xi^bh^{md}\\
        &=(\xi^f\xi^h\phi_{fh})\phi_e\phi_m R_{cbda}\xi^a\xi^bh^{md}v^cv^e-(h^{f}_c\xi^h\phi_{fh})\phi_e\phi_m R^c_{~bda}\xi^a\xi^bh^{md}v^e\\
        &~~~-(h^{h}_{e}\xi^f\phi_{fh})\phi^e\phi_m R^c_{~bda}\xi^a\xi^bh^{md}v^c+h^{he}\phi_{fh}\phi_e\phi_m (R_{cbda}\xi^b\xi^a\mathbf{e}^c_C\mathbf{e}^d_A)\mathbf{e}^m_B\mathbf{e}^f_D\delta^{AB}\delta^{CD}\\
        &=0~,
    \end{aligned}
\end{equation}
upon using the identities of Riemann tensor on $\mathcal{H}$ and the symmetry identities on $\phi$. Thus, the second term vanishes, and the expansion of $\nabla^c\mathcal{G}_5\phi^d R_{cabd}\xi^a\xi^b$ becomes trivial, so $\mathfrak{F}$ vanishes.

\section{Trivial part of the contraction  $\mathfrak{H}^{(5)}\xi^ah^b_e$~i.e. $\mathfrak{Z}_e$}\label{B}
In this appendix, we show that the terms independent of the curvature tensor in the mixed projection vanish when contracted with $\xi^ah^b_e$. The details of the algebraic steps are given below, where we use the previously obtained identities starting from $\pounds_{\xi}\phi=0$:
\begin{enumerate}
    \item \underline{$\mathfrak{I}^{(1)}$}: 
   The term $\mathfrak{I}^{(1)}\phi_{ab}\xi^ah^b_e$ vanishes identically due to \ref{eq19}.
   \item \underline{$\mathfrak{J}^{(1)}$}: 
   $\mathfrak{J}^{(1)}\phi_{a}\phi_b\xi^ah^b_e$ identically vanishes since $\phi_a\xi^a=0$.
   \item \underline{$\mathfrak{J}^{(2)}$}: We have $\phi_{ca}\phi^c_{~b}\xi^ah^b_e$ which can be simplified as 
   \begin{equation*}
   \begin{aligned}
       \phi_{ca}\phi^c_{~b}\xi^ah^b_e&= \phi_{ca}\phi_{db}g^{cd}\xi^ah^b_e=\phi_{ca}\phi_{db}(-\xi^cv^d-\xi^dv^c+h^{cd})\xi^ah^b_e\\
       &=-(\phi_{ca}\xi^c\xi^a)v^d\phi_{db}h^b_e-(\phi_{db}\xi^dh^b_e)v^c\phi_{ca}\xi^a+(\phi_{ca}\xi^ah^c_d)\phi^{d}_{~b}h^b_e=0~.
   \end{aligned}
   \end{equation*}
    \item  \underline{$\mathfrak{K}^{(1)}$}: The contraction $\phi^{c}\phi_{ca}\phi_b\xi^ah^b_e$ can be expanded out as:
    \begin{equation*}
        \begin{aligned}
            \phi^{c}\phi_{ca}\phi_b\xi^ah^b_e&=\phi_{d}g^{cd}\phi_{ca}\phi_b\xi^ah^b_e=\phi_{d}(-\xi^cv^d+h^{cd})\phi_{ca}\phi_b\xi^ah^b_e\\
            &=-(\phi_{ca}\xi^a\xi^c)\phi_b\phi_dv^dh^b_e+(\phi_{ca}\xi^ah^c_d)\phi^d\phi_bh^b_e=0~.
        \end{aligned}
    \end{equation*}
    \item  \underline{$\mathfrak{K}^{(2)}$}: The term $\phi_{cd}\phi^{c}_{~a}\phi^d_{~b}\xi^ah^b_e$ can be expanded as:
    \begin{equation*}
        \begin{aligned}
            \phi_{cd}\phi^{c}_{~a}\phi^d_{~b}\xi^ah^b_e&=\phi_{cd}\phi_{ma}\phi_{nb}g^{cm}g^{dn}\xi^ah^b_e\\&=\phi_{cd}\phi_{ma}\phi_{nb}(-\xi^cv^m-\xi^mv^c+h^{cm})(-\xi^dv^n-\xi^nv^d+h^{dn})\xi^ah^b_e\\
            &=-\phi_{cd}\phi_{ma}\phi_{nb}\xi^cv^m(-\xi^dv^n-\xi^nv^d+h^{dn})\xi^ah^b_e\\
            &=(\phi_{cd}\xi^c\xi^d)\phi_{ma}\phi_{nb}v^mv^n\xi^ah^b_e+(\phi_{nb}\xi^nh^b_e)\phi_{cd}\phi_{ma}v^mv^d\xi^a\xi^c\\
            &~~-(\phi_{cd}\xi^ch^d_n)\phi_{ma}\phi^n_{~b}v^m\xi^ah^b_e=0~.
        \end{aligned}
    \end{equation*}
    \item \underline{$\mathfrak{L}^{(1)}$}: For $\phi_{cd}\phi^d\phi^c_{~a}\phi_b\xi^ah^b_e$, we can look at the following expansion:
    \begin{equation*}
        \begin{aligned}
            \phi_{cd}\phi^d\phi^c_{~a}\phi_b\xi^ah^b_e&=\phi_{cd}\phi_m\phi_{na}(-\xi^dv^m+h^{dm})(-\xi^cv^n-\xi^nv^c+h^{cn})\phi_b\xi^ah^b_e\\
            &=\phi_{cd}\phi_m\phi_{na}(\xi^dv^m-h^{dm})\xi^cv^n\phi_b\xi^ah^b_e\\
            &=(\phi_{cd}\xi^c\xi^d)\phi_m\phi_{na}v^mv^n\phi_b\xi^ah^b_e-(\phi_{cd}\xi^ch^d_m)\phi^m\phi_{na}v^n\phi_b\xi^ah^b_e=0~.
        \end{aligned}
    \end{equation*}
    \item \underline{$\mathfrak{L}^{(2)}$}: Clearly $\phi^c\phi^d\phi_{ca}\phi_{db}\xi^ah^b_e$ vanishes as
    \begin{equation*}
        \begin{aligned}
            \phi^c\phi^d\phi_{ca}\phi_{db}\xi^ah^b_e&=\phi_m\phi_n(-\xi^cv^m+h^{cm})(-\xi^dv^n+h^{dn})\phi_{ca}\phi_{db}\xi^ah^b_e\\
            &=\phi_m\phi_nh^{cm}(-\xi^dv^n+h^{dn})\phi_{ca}\phi_{db}\xi^ah^b_e\\
            &=-(\phi_{db}\xi^dh^b_e)v^n\phi_m\phi_nh^{cm}\phi_{ca}\xi^a+(\phi_{ca}\xi^ah^{c}_m)\phi^m\phi_nh^{dn}\phi_{db}h^b_e=0~.
        \end{aligned}
    \end{equation*}
    \item \underline{$\mathfrak{M}^{(1)}$}: We have the expression $\phi^{cf}\phi_f\phi^d\phi_{ca}\phi_{db}\xi^ah^b_e$ which yields
    \begin{equation*}
        \begin{aligned}
            \phi^{cf}\phi_f\phi^d\phi_{ca}\phi_{db}\xi^ah^b_e&=\phi_{mn}(-\xi^cv^m-\xi^mv^c+h^{cm})(-\xi^nv^f+h^{nf})\phi_f\phi_p(-\xi^dv^p+h^{dp})\phi_{ca}\phi_{db}\xi^ah^b_e\\
            &=\phi_{mn}\xi^mv^c(\xi^nv^f-h^{nf})\phi_f\phi_p(-\xi^dv^p+h^{dp})\phi_{ca}\phi_{db}\xi^ah^b_e\\
            &=(\phi_{mn}\xi^m\xi^n)v^cv^f\phi_f\phi_p(-\xi^dv^p+h^{dp})\phi_{ca}\phi_{db}\xi^ah^b_e\\
            &~~-(\phi_{mn}\xi^mh^n_f)v^c\phi^f\phi_p(-\xi^dv^p+h^{dp})\phi_{ca}\phi_{db}\xi^ah^b_e=0~.\\
        \end{aligned}
    \end{equation*}
    \item \underline{$\mathfrak{M}^{(2)}$}: The term $\mathfrak{M}^{(2)}\phi^{f}\phi_{fd}\phi^{cd}\phi_{ca}\phi_b\xi^ah^b_e$ goes to zero since we have
    \begin{equation*}
        \begin{aligned}
            \phi^{f}\phi_{fd}\phi^{cd}\phi_{ca}\phi_b\xi^ah^b_e&=(-\xi^fv^m+h^{fm})\phi_m\phi_{fd}\phi_{np}(-\xi^cv^n-\xi^nv^c+h^{cn})(-\xi^dv^p-\xi^pv^d+h^{dp})\phi_{ca}\phi_b\xi^ah^b_e\\
            &=(\xi^fv^m-h^{fm})\phi_m\phi_{fd}\phi_{np}\xi^nv^c(-\xi^dv^p-\xi^pv^d+h^{dp})\phi_{ca}\phi_b\xi^ah^b_e\\
            &=-(\xi^fv^m-h^{fm})\phi_m\phi_{fd}\phi_{np}\xi^nv^c\xi^dv^p\phi_{ca}\phi_b\xi^ah^b_e\\
            &=-(\phi_{fd}\xi^f\xi^d)\phi_mv^m\phi_{np}\xi^nv^cv^p\phi_{ca}\phi_b\xi^ah^b_e+(\phi_{fd}h^f_m\xi^d)\phi^m\phi_{np}\xi^nv^cv^p\phi_{ca}\phi_b\xi^ah^b_e\\
            &=0~.
        \end{aligned}
    \end{equation*}
    \item \underline{$\mathfrak{M}^{(3)}$} : For $\phi_{cf}\phi^c\phi^f\phi_{da}\phi^d_{~b}\xi^ah^b_e$, we have:
    \begin{equation*}
        \begin{aligned}
            \phi_{cf}\phi^c\phi^f\phi_{da}\phi^d_{~b}\xi^ah^b_e&=\phi_{cf}\phi_m(-\xi^cv^m+h^{cm})\phi_n(-\xi^fv^n+h^{fn})\phi_{da}\phi_{pb}(-\xi^pv^d-\xi^dv^p+h^{pd})\xi^ah^b_e\\
            &=\phi_{cf}\phi_m(\xi^cv^m-h^{cm})\phi_n(-\xi^fv^n+h^{fn})\phi_{da}v^d\xi^a(\phi_{pb}\xi^ph^b_e)=0~.\\
        \end{aligned}
    \end{equation*}
 
    \item \underline{$\mathfrak{M}^{(4)}$} : And lastly, the contraction $\phi^{cf}\phi_f\phi^d\phi_{da}\phi_{cb}\xi^ah^b_e$ identically vanishes as
    \begin{equation*}
        \begin{aligned}
            \phi^{cf}\phi_f\phi^d\phi_{da}\phi_{cb}\xi^ah^b_e&=\phi_{mn}(-\xi^mv^c+h^{cm})(-\xi^nv^f+h^{nf})\phi_f\phi_p(-\xi^dv^p+h^{dp})\phi_{cb}\phi_{da}\xi^ah^b_e\\
            &=0~.
        \end{aligned}
    \end{equation*}
\end{enumerate}
Hence, the contraction $\mathfrak{H}^{(5)}\xi^ah^b_e$ yields only terms dependent on the curvature tensor $\mathfrak{R}_e$, and the rest i.e. $\mathfrak{Z}_e$ is trivial.
\section{Simplified expression of $\mathfrak{R}_e$}\label{C}
Before we go over the details, we shall provide two relations: \\
For a differentiable scalar function $f$ defined over $\mathscr{M}$, we have the following\footnote[4]{Since, $\nabla^mf=D^mf-\xi^m(\nabla^v_{\perp}f)-v^m(\nabla^{\xi}_{\perp}f)$, where $\nabla^{\xi}_{\perp}\equiv \xi^n\nabla_n f$ and similarly $\nabla^{v}_{\perp}\equiv v^n\nabla_n f$.} 
\begin{equation}
    (\phi^m)(D_m f)=(D^m \phi)(D_m f)~,
\end{equation}
and 
\begin{equation}
   (D_m f) \phi^{mn}(D_n \phi)=(D_m f) (D^mD^n \phi)(D_n \phi)~.
\end{equation}
\subsection{Vanishing terms}
We show that certain terms in $\mathfrak{R}_e$ vanish due to the symmetry properties of $R_{abcd}$, the identities \ref{geom_identities} on $\mathcal{H}$, and the fact that $\phi$ is Lie dragged.\\
In \ref{curv}, let us consider the term $(\mathcal{G}_{4X}-\frac{1}{2}\mathcal{G}_{5X}\Box\phi)R_{acbd}\phi^c\phi^d\xi^ah^b_e$, where we find that
\begin{equation*}
    \begin{aligned}
        R_{acbd}\phi^c\phi^d\xi^ah^b_e&=R_{acbd}\phi_m\phi_n(-\xi^cv^m+h^{cm})(-\xi^dv^n+h^{dn})\xi^ah^b_e=R_{acbd}\phi_m\phi_nh^{cm}(-\xi^dv^n+h^{dn})\xi^ah^b_e\\
        &=(R_{acdb}\xi^a\mathbf{e}^c_A\xi^d\mathbf{e}^b_C)\phi_m\phi_nv^n\mathbf{e}^m_B\mathbf{e}_e^D\delta^{AB}\delta^C_D-(R_{acdb}\xi^a\mathbf{e}^{c}_A\mathbf{e}^{d}_C\mathbf{e}^b_E)\mathbf{e}^{m}_B\mathbf{e}^{n}_D\mathbf{e}_e^F\phi_m\phi_n\delta^{AB}\delta^{CD}\delta^E_F\\
        &=0~,
    \end{aligned}
\end{equation*}
indicating that the first term in $\mathfrak{R}_e$ vanishes.\\
The next two terms that vanish are $\nabla^{c}\mathcal{G}_5\phi^dR_{c(ab)d}\xi^ah^b_e$, and $\mathcal{G}_{5X}R_{cdf(a}\phi^{~d}_{b)}\phi^c\phi^f\xi^ah^b_e$, since firstly, it is straightforward to show that both
\begin{equation*}
    \begin{aligned}
        \phi^c\phi^dR_{c(ab)d}\xi^ah^b_e=0~;~\phi^{cf}\phi_fR_{c(ab)d}\xi^ah^b_e=0
    \end{aligned}
\end{equation*}
and secondly,
\begin{equation*}
    \begin{aligned}
        R_{cdfa}\phi^{~d}_{b}\phi^c\phi^f\xi^ah^b_e&=R_{cdfa}\phi_{bm}\phi_n\phi_p(-\xi^dv^m+h^{md})(-\xi^cv^n+h^{cn})h^{fp}\xi^ah^b_e\\
        &=R_{cdfa}\phi_{bm}\phi_n\phi_ph^{md}(-\xi^cv^n+h^{cn})h^{fp}\xi^ah^b_e=0~,
    \end{aligned}
\end{equation*}
and 
\begin{equation*}
    \begin{aligned}
        R_{cdfb}\phi^{~d}_{a}\phi^c\phi^f\xi^ah^b_e&=R_{cdfb}\phi_{am}\phi_n\phi_p\xi^dv^mh^{cn}(\xi^fv^p-h^{fp})\xi^ah^b_e=0~.\\
    \end{aligned}
\end{equation*}
These are the only terms that vanish due to the properties of the Riemann tensor.
\subsection{Non-vanishing terms}
Since the expression for $\mathfrak{R}_e$ does not identically vanish, there exist expressions that one can cast in terms of $D_e\kappa$ by careful algebra, starting with
$(\mathcal{G}_{4X}-\frac{1}{2}\mathcal{G}_{5X}\Box\phi)R_{ca}\phi_{b}\phi^c\xi^ah^b_e$ which becomes
$-(\mathcal{G}_{4X}-\frac{1}{2}\mathcal{G}_{5X}\Box\phi)(D^d\phi)(D_d\kappa)(D_e\phi)$, as
\begin{equation}\label{nv-1}
    \begin{aligned}
        R_{ca}\phi_{b}\phi^c\xi^ah^b_e&=\mathcal{G}_{4X}R_{ca}\phi_{b}\phi_d(-\xi^cv^d+h^{dc})\xi^ah^b_e\\&=-(R_{ca}\xi^c\xi^a)\phi_{b}\phi_dv^dh^b_e+(R_{ca}\xi^ah^{c}_d)\phi_{b}\phi^dh^b_e\\
        &=-(D^d\phi)(D_d\kappa)(D_e\phi)~.
    \end{aligned}
\end{equation}
Next, the expression $\frac{1}{2}\mathcal{G}_{5X}R_{cd}\phi^{c}\phi^d_{~a}\phi_{b}\xi^ah^b_e$ can be rewritten in terms of $\kappa$ and $\phi$ by noting that
\begin{equation}
    \begin{aligned}
        R_{cd}\phi^{c}\phi^d_{~a}\phi_{b}\xi^ah^b_e&=R_{cd}\phi_{m}(-\xi^cv^m+h^{cm})\phi_{na}(-\xi^nv^d-\xi^dv^n+h^{dn})\phi_{b}\xi^ah^b_e\\
        &=R_{cd}\phi_{m}(\xi^cv^m-h^{cm})\phi_{na}\xi^dv^n\phi_{b}\xi^ah^b_e\\
        &=-(R_{cd}h^c_m\xi^d)\phi^m(\phi_{na}v^n\xi^a)\phi_bh^b_e=(D^m\phi)(D_m\kappa)(D_e\phi)\Phi
    \end{aligned}
\end{equation}
with $\Phi\equiv \phi_{np}v^n\xi^p$~. Thus, we have
\begin{equation}\label{nv-2}
    \frac{1}{2}\mathcal{G}_{5X}R_{cd}\phi^{c}\phi^d_{~a}\phi_{b}\xi^ah^b_e=\frac{1}{2}\mathcal{G}_{5X}(D^m\phi)(D_m\kappa)(D_e\phi)\Phi~.
\end{equation}
Now, we consider the expression $\frac{1}{2}\mathcal{G}_{5X}R_{cdfa}\phi_{b}\phi^d\phi^{cf}\xi^ah^b_e$ where, we note that
\begin{equation*}
    \begin{aligned}
        R_{cdfa}\phi_{b}\phi^d\phi^{cf}\xi^ah^b_e&=R_{cdfa}\phi_{b}\phi_m(-\xi^dv^m+h^{dm})\phi_{np}(-\xi^cv^n-\xi^nv^c+h^{cn})(-\xi^pv^f+h^{pf})\xi^ah^b_e\\
        &=R_{cdfa}\phi_{b}\phi_m\xi^dv^m\phi_{np}(\xi^nv^c-h^{cn})(-\xi^pv^f+h^{pf})\xi^ah^b_e\\
        &~+R_{cdfa}\phi_{b}\phi_mh^{dm}\phi_{np}(-\xi^cv^n+h^{cn})(-\xi^pv^f+h^{pf})\xi^ah^b_e\\
        &=-(R_{cdfa}h^{c}_n\xi^dh^{f}_p\xi^a)\phi_{b}\phi_mv^m\phi^{np}h^b_e+R_{cdfa}\phi_{b}\phi_mh^{dm}\phi_{np}(-\xi^cv^n+h^{cn})(-\xi^pv^f+h^{pf})\xi^ah^b_e\\
        &=R_{cdfa}\phi_{b}\phi_mh^{dm}\phi_{np}(-\xi^cv^n+h^{cn})(-\xi^pv^f+h^{pf})\xi^ah^b_e\\
        &=R_{cdfa}\phi_{b}\phi_mh^{dm}\phi_{np}\xi^cv^n\xi^pv^f\xi^ah^b_e=(R_{afdc}\xi^av^fh^{d}_m\xi^c)\phi^m(\phi_{np}v^n\xi^p)\phi_bh^b_e~,
    \end{aligned}
\end{equation*}
and since, $R_{afdc}\xi^av^fh^{d}_m\xi^c=D_m \kappa$, it immediately follows that
\begin{equation}
    \frac{1}{2}\mathcal{G}_{5X}R_{cdfa}\phi_{b}\phi^d\phi^{cf}\xi^ah^b_e=\frac{1}{2}\mathcal{G}_{5X}\phi^m(D_m \kappa) \Phi D_{e}\phi=\frac{1}{2}\mathcal{G}_{5X}(D^m\phi)(D_m \kappa)(D_{e}\phi)\Phi~.
\end{equation}
Lastly, we analyze the final two expressions in $\mathfrak{R}_e$ namely: $-\frac{1}{2}\nabla^c\mathcal{G}_5R_{ca}\phi_{b}\xi^ah^b_e~,$ and $\nabla_{(a}\mathcal{G}_5 G_{b)c}\phi^c\xi^ah^b_e$.
First, using \ref{nv-1}, we get
\begin{equation}
    \begin{aligned}
        -\frac{1}{2}\mathcal{G}_{5\phi}R_{ca}\phi^c\phi_b\xi^ah^b_e=\frac{1}{2}\mathcal{G}_{5\phi}(D^d\phi)(D_d\kappa)(D_e\phi)~.
    \end{aligned}
\end{equation}
Moreover, 
\begin{equation}
    \begin{aligned}
        \phi^{cd}\phi_dR_{ca}\phi_b\xi^ah^b_e&=\phi_{mn}\phi_dh^{cm}h^{nd}R_{ca}\phi_b\xi^ah^b_e\\
        &=(R_{ca}h^c_m\xi^a)\phi^{m}_{~n}h^{nd}\phi_d\phi_bh^b_e=-(D_m\kappa)(D_n\phi)(D_e\phi)\phi^{mn}~.
    \end{aligned}
\end{equation}
Hence, we have 
\begin{equation}
    -\frac{1}{2}\nabla^c\mathcal{G}_5R_{ca}\phi_{b}\xi^ah^b_e=\frac{1}{2}\mathcal{G}_{5\phi}(D^d\phi)(D_d\kappa)(D_e\phi)-\frac{1}{2}\mathcal{G}_{5X}(D_m\kappa)(D_n\phi)(D_e\phi)(D^m D^n \phi)~.
\end{equation}
Finally, for the term $\nabla_{(a}\mathcal{G}_5 G_{b)c}\phi^c\xi^ah^b_e$ it is straightforward to show that
\begin{equation}
    \nabla_{a}\mathcal{G}_5 G_{bc}\phi^c\xi^ah^b_e=0~.
\end{equation}
However, $\nabla_{b}\mathcal{G}_5 G_{ac}\phi^c\xi^ah^b_e\ne0$, as
\begin{equation}
    \phi_b G_{ac}\phi^c\xi^ah^b_e=\phi_b\phi^m (R_{ac}\xi^ah^{c}_m)h^b_e=-(D^m\phi)(D_m\kappa)(D_e\phi)
\end{equation}
and
\begin{equation}
    \phi_{bd}\phi^dG_{ac}\phi^c\xi^ah^b_e=\phi_{bd}\phi_f\phi^mh^{df}(R_{ca}h^c_m\xi^a)h^b_e=-(D^m\phi)(D_m\kappa)(D_eD_n\phi)(D^n\phi)~,
\end{equation}
thereby yielding the following reduction
\begin{equation}
    \nabla_{(a}\mathcal{G}_5 G_{b)c}\phi^c\xi^ah^b_e=-\frac{1}{2}\mathcal{G}_{5\phi}(D^m\phi)(D_m\kappa)(D_e\phi)+\frac{1}{2}\mathcal{G}_{5X}(D^m\phi)(D_m\kappa)(D_eD_n\phi)(D^n\phi)~.
\end{equation}
Thus, we arrive at the final expression for $\mathfrak{R}_e$ as follows
\begin{equation}
\begin{aligned}
    \mathfrak{R}_e&=-(\mathcal{G}_{4X}-\frac{1}{2}\mathcal{G}_{5X}\Box\phi)(D^m\phi)(D_m\kappa)(D_e\phi)+\mathcal{G}_{5X}(D^m\phi)(D_m\kappa)(D_e\phi)\Phi~\\
    &~+\mathcal{G}_{5\phi}(D^m\phi)(D_m\kappa)(D_e\phi)-\frac{1}{2}\mathcal{G}_{5X}(D_m\kappa)(D_n\phi)(D_e\phi)(D^m D^n \phi)\\
    &~-\frac{1}{2}\mathcal{G}_{5X}(D^m\phi)(D_m\kappa)(D_eD_n\phi)(D^n\phi)~.
\end{aligned}
\end{equation}
The above simplications have been incorporated in the main text.


\bibliographystyle{utphys1}
\bibliography{mybib}

\end{document}